\documentclass[12pt]{article}
\usepackage{a4wide}
\usepackage{amssymb}
\usepackage{graphicx}
\begin{document}
{\renewcommand{\thefootnote}{\fnsymbol{footnote}}
\begin{center}
{\LARGE  Hypersurface deformations}\\
\vspace{1.5em}
Martin Bojowald,\footnote{e-mail address: {\tt bojowald@psu.edu}}
Erick I.~Duque\footnote{e-mail address: {\tt eqd5272@psu.edu}}
and Aiden Shah\footnote{e-mail address: {\tt aidenshah@outlook.com}}
\\
\vspace{0.5em}
Institute for Gravitation and the Cosmos,\\
The Pennsylvania State
University,\\
104 Davey Lab, University Park, PA 16802, USA\\
\vspace{1.5em}
\end{center}
}

\setcounter{footnote}{0}

\begin{abstract}
  Deformations of spacelike hypersurfaces in space-time play an important role
  in discussions of general covariance and slicing independence in
  gravitational theories. In a canonical formulation, they provide the
  geometrical meaning of gauge transformations generated by the diffeomorphism
  and Hamiltonian constraints. However, it has been known for some time that
  the relationship between hypersurface deformations and general covariance is
  not a kinematical equivalence but holds only on the solution space of the
  constraints and requires their gauge equations and equations of motion to
  be used. The off-shell behavior of hypersurface deformations on their own,
  without imposing constraint and gauge equations, is therefore different from
  space-time diffeomorphisms. Its complete understanding is important for
  potential quantizations or modifications of general relativity in canonical
  form and of compatible space-time geometries that may be implied by
  them. Here, a geometrical analysis of hypersurface deformations is
  performed, allowing for a dependence of hypersurface deformation generators
  (the lapse function and the shift vector) on the phase-space degrees of
  freedom given by the geometry of an embedded spacelike hypersurface. The
  result is compared in detail with Poisson brackets of the gravitational
  constraints.  As a new implication of physical relevance, covariance
  conditions are obtained for theories of emergent modified gravity without
  symmetry restrictions.
\end{abstract}

\section{Introduction}

In canonical gravity, time evolution and gauge transformations are generated
by the gravitational constraints via Poisson brackets. The geometrical picture
of these transformations describes infinitesimal deformations of an initial
spacelike hypersurface in an embedding space-time, combined with possible
transformations within the hypersurface. Since there is no distinguished time
direction in covariant space-time, the only intrinsic notion for a deformation
transversal to the hypersurface is the normal direction, given by the
future-pointing unit normal $n^{\mu}$.

Accordingly, the four dimensions of a space-time direction are broken up into
transformations tangential to the hypersurface (generated by the
diffeomorphism constraint $\vec{H}[\vec{M}]$ with a spatial shift vector
$\vec{M}$ tangential to the hypersurface) and a transformation normal to the
hypersurface (generated by the Hamiltonian constraint $H[N]$ with a normal
displacement by the lapse function $N$ in the direction $n^{\mu}$). These
transformations do not have a direct relationship with space-time
diffeomorphisms or coordinate changes because the latter are expressed in a
frame that makes use of a time direction rather than normals to spacelike
hypersurfaces. A correspondence between hypersurface deformations and
space-time diffeomorphisms can therefore be achieved only on-shell, using
equations of motion generated by the constraints in order to introduce a time
direction. The covariant freedom of choosing the time direction is expressed
by hypersurface deformations through the free gauge functions, lapse $N$ and
shift $\vec{M}$.

Equations of motion for an on-shell correspondence require explicit
phase-space functions for the constraints, rather than abstract algebraic
generators. Gauge transformations and equations of motion can then be derived
as Hamilton's equations, generated by a combination $H[N]+\vec{H}[\vec{M}]$ of
the constraint expressions, where $N$ and $\vec{M}$ may be used to define a
time-evolution vector field in space-time, or a vector field that may be
compared to an infinitesimal space-time diffeomorphism. The phase-space, as
derived from a canonical analysis of general relativity or some alternative
formulation \cite{DiracHamGR,Katz,ADM}, can be expressed by configuration
variables that determine the spatial metric $q_{ab}$ of a hypersurface
together with momenta $p^{ab}$ whose geometrical meaning (classically, related
to extrinsic curvature of the hypersurface) follows from the equations of
motion for $q_{ab}$ generated by the resulting constraints.

The $3+1$ independent transformations generated by $H[N]+\vec{H}[\vec{M}]$ are
brought back into a 4-dimensional structure (per spatial point) by requiring
that Poisson brackets of the constraints have a suitable form, given by
\begin{eqnarray}
    \{ \vec{H} [ \vec{N}] , \vec{H} [ \vec{M} ] \} &=& - \vec{H}
                                                       [\mathcal{L}_{\vec{M}}
                                                       \vec{N}] 
    \ ,  \label{DD}\\
    \{ H [ N ] , \vec{H} [ \vec{M}]\} &=& - H [ M^b \partial_b N ]
    \ , \label{HD}\\
    \{ H [ N ] , H [ M ] \} &=& \vec{H} [ \sigma q^{a b} \left( M
                                \partial_b N - N \partial_b M \right)] \label{HH}
\end{eqnarray}
where, for the sake of completeness, we consider embeddings in Lorentzian
space-time ($\sigma=-1$) as well as 4-dimensional space of Euclidean signature
($\sigma=1$). The relationship between these Poisson brackets, derived from a
canonical formulation of general relativity, and the geometry of hypersurface
deformations has been derived in \cite{Regained,LagrangianRegained}. 

The brackets have the characteristic and complicating feature of structure
functions, depending not only on the generators $N$ and $\vec{M}$ but also on
the inverse spatial metric $q^{ab}$ induced on a spacelike hypersurface. An
open question is what general mathematical structure (generalizing a Lie
algebra) might be suitable for off-shell hypersurface deformations. Older
attempts tried to define the ``Bergmann--Komar group''
\cite{BergmannKomarGroup} by formal exponentiation of the
hypersurface-deformation generators. Such a construction, exported from the
simpler case of a Lie algebra, amounts to iterating Poisson brackets with
$H[N]$ as well as $\vec{H}[\vec{N}]$ as a formal power series modeled on the
Taylor expansion of the exponential function. If the computation of Poisson
brackets is iterated, as in $\{H[N_1],\{H[N_2],H[N_3]\}\}$ and so on, the
gauge functions inserted in the constraints, which in the initial derivation
of the constraint brackets (\ref{DD})--(\ref{HH}) depend only on spatial
coordinates on a given hypersurface and also on time if a family of
hypersurfaces is considered, become metric dependent as a consequence of the
structure function $q^{ab}$. But if $N$ and $\vec{N}$ are metric dependent,
the Poisson brackets (\ref{DD})--(\ref{HH}) obtain additional terms with
partial derivatives of these functions by the metric components. The extended
brackets, shown in \cite{BergmannKomarGroup}, rather than the original ones,
should therefore be used for a self-consistent ``exponentiation'' of the
generators. The relationship between such extra terms and geometrical
properties of hypersurface deformations will be analyzed here in an off-shell
setting.

More recently, the papers \cite{ConsAlgebroid} and \cite{ConsRinehart}
constructed first a Lie algebroid that captures the on-shell behavior of
hypersurface-deformation brackets and agrees with (\ref{DD})--(\ref{HH}) for
metric-independent $N$ and $\vec{N}$, and then an $L_{\infty}$-algebroid for
the general case. The algebroid picture was motivated by the presence of
structure functions, while the $L_{\infty}$-bracket in the general case
followed from a formulation of the gauge transformations of general relativity
in a BRST/BFV formulation adapted to theories with structure functions. These
results describe algebraic properties of gauge transformations, motivated by a
certain gauge structure suitable for a perturbative treatment of interactions
as in the original BRST/BFV theories. While these algebraic properties are
important in a perturbative context, they do not directly reveal the geometry
of hypersurface deformations in space-time.

A puzzling feature is given by the result that an
$L_{\infty}$-bracket is in general necessary, which does not obey the Jacobi
identity in contrast to the Poisson bracket used in the original derivation of
(\ref{DD})--(\ref{HH}) or in extensions to phase-space dependent lapse
functions and shift vectors. Moreover, the intuitive understanding of
infinitesimal hypersurface deformations, which as deformations in an embedding
space-time are associative, shows no hint as to why an $L_{\infty}$-bracket
should appear in their algebraic description. There are therefore three
formulations of hypersurface deformations: the algebraic one with
$L_{\infty}$-algebroids (in which the metric is accompanied by the ghost and
antifields of a BRST/BFV formulation), the geometrical one using direct
embeddings in space-time, and the canonical one based on Poisson brackets
(\ref{DD})--(\ref{HH}). The first one does not agree with the latter two, and
the relationship between these latter two has not been fully explored,
suggesting several questions that motivate the present analysis. Such an
analysis is of practical importance in canonical quantum gravity because an
appearance of non-associative structures or deviations between the canonical
and geometrical pictures could further complicate this already challenging
approach. As we will see, there are also classical implications in the context
of modified gravity theories.

The inevitability of metric-dependent $N$ and $\vec{N}$ in iterated Poisson
brackets of the constraints is a direct implication of the structure function
in a phase-space representation of the constraints and their brackets. This
property requires an extension of the geometrical picture of hypersurface
deformations if the direction of a deformation depends not only on the
position on a hypersurface but also on its local induced geometry. More
generally, this dependence may also be extended to the momentum canonically
conjugate to the spatial metric, which in classical general relativity is
related to the extrinsic curvature of a hypersurface in space-time. As we will
see, such dependencies are not only of mathematical interest, but also have
unexpected physical implications in the construction of new generally
covariant modified gravity theories \cite{Higher,HigherCov,SphSymmMinCoup}, or
in geometrical interpretations of possible quantum effects implied by
canonical gravity. These theories of emergent modified gravity maintain the
geometrical structure of hypersurface deformations but realize them in
space-times with modified dynamics, compared with general relativity. The aim
of this paper is therefore to provide a complete analysis of the geometry of
hypersurface deformations in which deformation directions may depend on the
local intrinsic and extrinsic geometry. This analysis will allow us to derive
new conditions on possible theories of emergent modified gravity that, unlike
previous approaches, do not require symmetry assumptions on space-time
solutions.

Section~\ref{s:Standard} provides a brief review of the standard ingredients
of canonical gravity, focusing on how gauge transformations are related to
coordinate changes and general covariance. In Section~\ref{sec:Geometrodynamics revised}, we
perform a detailed geometrical analysis of hypersurface deformations including
off-shell properties as well as phase-space dependent gauge functions,
followed in Section~\ref{s:Offshell} by a corresponding
canonical analysis. In Sections~\ref{s:Complete} and \ref{s:Dependent}, we conclude
that the geometrical and canonical sides of hypersurface deformations are not
equivalent to each other. After an application of our results to the
construction of new gravitational theories of emergent modified gravity in
Section~\ref{s:new}, implications for canonical quantum gravity are discussed
in our final Section~\ref{s:QG} before our conclusions, which will relate our
results to foundational questions in canonical quantum gravity.

\section{Standard canonical theory}
\label{s:Standard}

In order to set up the canonical formulation, we assume that space-time (or
4-dimensional space with Euclidean signature) has a globally hyperbolic topology,
$M = \Sigma \times \mathbb{R}$ with a 3-dimensional manifold $\Sigma$, and a
geometry defined by the line element
\begin{equation}
    {\rm d} s^2 = \sigma N^2 {\rm d} t^2 + q_{a b} ( {\rm d} x^a + N^a {\rm d}
    t ) ( {\rm d} x^b + N^b {\rm d} t )
    \,.
    \label{eq:ADM line element}
\end{equation}
The time direction in the given frame of (\ref{eq:ADM line element}) is
described by the time-evolution vector field
\begin{equation}
    t^\mu = N n^\mu + N^a s_a^\mu
    \label{eq:Time-evolution vector field}
\end{equation}
where $n^\mu$ is the unit normal to $\Sigma_t=\{X\in M\colon t(X)=t\}$, and
$s_a^\mu$ are the remaining three basis vectors tangential to the hypersurface
$\Sigma$. Therefore, $g_{\mu \nu} n^\mu s^\mu_a = 0$. The components $N$ and
$N^a$ in (\ref{eq:ADM line element}) and (\ref{eq:Time-evolution vector
  field}), respectively, are the same because the inverse space-time metric
and spatial metric are related by
$g^{\mu\nu}=q^{\mu\nu}+\sigma n^{\mu}n^{\nu}$. Replacing $n^{\mu}$ with $t^{\mu}$
using (\ref{eq:Time-evolution vector field}) and inverting the resulting
tensors then implies that $g_{\mu\nu}{\rm d}x^{\mu}{\rm d}x^{\nu}$ equals
(\ref{eq:ADM line element}).  As usual, $N$ --- the projection of $t^\mu$
normal to $\Sigma_t$ --- is called the lapse function, and $N^a$ --- the
projection of $t^\mu$ onto $\Sigma_t$ --- is called the shift vector.  Here,
we use Greek letters for spacetime indices (on $TM$), and latin letters for
spatial indices (on $T\Sigma$).

In this formulation, the Hamiltonian of general relativity is completely
constrained, $H[N] + H_a[N^a] = 0$ for all $N$ and $N^a$, where $H$ is the
Hamiltonian constraint and $H_a$ the diffeomorphism constraint. The lapse
function and shift vector appear as Lagrange multipliers in a direct
derivation of the constraints by a Legendre transformation of the
Einstein--Hilbert action. For phase-space independent $N$ and $N^a$, the
constraints obey the brackets (\ref{DD})--(\ref{HH}) and are therefore first
class, generating gauge transformations that should in some way correspond to
the gauge content of coordinate transformations realized in manifest form in
the usual formulation of general relativity based on space-time tensors. This
correspondence requires several steps to be fully realized.

Given a theory with constraints, the dynamical solution space is the subspace
of the kinematic space of initial-value configurations (given by the phase
space of spatial metrics and momenta on $\Sigma$) on which the constraints
vanish $H[N]=0=H_a[N^a]$ for all $x^a$-dependent $N$ and $N^a$. (We ignore
boundary considerations in this paper.) In addition, the constraints generate
gauge transformations for any phase space function $\mathcal{O}$ via the
Poisson bracket,
$\delta_\epsilon \mathcal{O} = \{ \mathcal{O} , H[\epsilon^0]+H_a[ \epsilon^a]
\}$. In a completely constrained theory in which the Hamiltonian is a
constraint, they also generate time evolution via Hamilton's equations of
motion:
$\dot{\mathcal{O}} \equiv \delta_t \mathcal{O} = \{ \mathcal{O} ,
H[N]+H_a[N^a] \}$ where $N$ and $N^a$ appear as components of the
time-evolution vector field (\ref{eq:Time-evolution vector field}) in the
normal frame.  Thus, the dynamics is encoded by the same constraints that
generate gauge transformations.

Poisson-bracket relations $\delta_{\epsilon}\mathcal{O}$ for gauge
transformations only determine how the spatial metric and its momentum
transform, which is not sufficient for a complete transformation of the
space-time line element (\ref{eq:ADM line element}). However, the consistency
condition that the time evolution of gauge-transformed initial values is equal
to a gauge transformation of the evolved initial values determines gauge
transformations also for the lapse function $N$ and shift vector $\vec{N}$
that appear in the time-evolution vector field, constituting the remaining
components of the space-time metric.  Using the Jacobi identity, these gauge
transformations are given by \cite{LapseGauge,CUP}
\begin{eqnarray}\label{eq:Off-shell gauge transformations for lapse}
    \delta_\epsilon N &=& \dot{\epsilon}^0 + \epsilon^a \partial_a N - N^a \partial_a \epsilon^0\\
    \delta_\epsilon N^a &=& \dot{\epsilon}^a + \epsilon^b \partial_b N^a - N^b \partial_b \epsilon^a
    - \sigma q^{a b} \left(\epsilon^0 \partial_b N - N \partial_b \epsilon^0 \right)
    \label{eq:Off-shell gauge transformations for shift}
\end{eqnarray}
with terms determined by the structure constants and structure functions of
(\ref{DD})--(\ref{HH}) as they appear in a commutator of gauge transformations
and time evolution.  These equations might change should the Jacobi identity
no longer hold for abstract versions of the generators, potentially
challenging the correspondence between gauge transformations and space-time
diffeomorphisms.

In the canonical formulation of general relativity, Poisson brackets with the
constraints encode general covariance in the following way.  Gauge
transformations of the spatial metric correspond to infinitesimal space-time
diffeomorphisms or Lie derivatives according to
\begin{eqnarray}
    \{ q_{a b} , \vec{H} [\vec{\epsilon}] \} \big|_{{\rm O.S.}}
    = \mathcal{L}_{\vec{\epsilon}} q_{a b} \quad,\quad
    \{ q_{a b} , H [\epsilon^0] \} \big|_{{\rm O.S.}}
    = \mathcal{L}_{\epsilon^0 n} q_{a b}\,.
    \label{eq:On-shell gauge transformation of 3-metric}
\end{eqnarray}
Here, ``O.S.'' (for on-shell) means that the constraints must be solved and
the gauge-transformation and evolution equations they generate must be used in
order to relate derivatives on both sides of these equations. If the equations
are not taken on-shell in this sense, gauge transformations generated by the
constraints are distinct from infinitesimal space-time diffeomorphisms.
On-shell, together with the gauge transformations of the lapse function and
shift vector, given by (\ref{eq:Off-shell gauge transformations for lapse})
and (\ref{eq:Off-shell gauge transformations for shift}), the transformations
reproduce infinitesimal diffeomorphisms of the full metric (\ref{eq:ADM line
  element}),
\begin{eqnarray}
    \delta_\epsilon g_{\mu \nu} \big|_{{\rm O.S.}} &=&
    \mathcal{L}_{\xi} g_{\mu \nu}\,.
    \label{eq:Covariance identity - classical GR}
\end{eqnarray}
The Lie derivative $\mathcal{L}_{\xi}$ is taken along the space-time vector
field $\xi$ related to the canonical gauge generators
$(\epsilon^0, \epsilon^a)$ by a linear transformation from a frame adjusted to
the foliation (using the unit normal $n^{\mu}$) to a coordinate frame (with
time direction $t^{\mu}$) \cite{BergmannKomarGroup}:
\begin{equation}\label{eq:Diffeomorphism generator projection}
    \xi^\mu = \epsilon^0 n^\mu + \epsilon^a s^\mu_a
    = \xi^t t^\mu + \xi^a s^\mu_a
  \end{equation}
  with components
  \begin{equation} \label{xieps}
    \xi^t = \frac{\epsilon^0}{N}
    \quad\mbox{and}\quad
    \xi^a = \epsilon^a - \frac{\epsilon^0}{N} N^a
    \,.
\end{equation}

\section{Geometry of hypersurface deformations}
\label{sec:Geometrodynamics revised}

We now turn our attention to the underlying geometrical picture of
hypersurface deformations, related to the Poisson brackets of the canonical
constraints. A comparison between geometrical and canonical formulations in
the following section will then reveal whether Poisson brackets of the
gravitational constraints provide a faithful off-shell description of
hypersurface deformations that can be realized in a 4-dimensional
geometry. The first two subsections relate properties of space-time Lie
derivatives to hypersurface deformations, revealing new off-shell features,
while Section~\ref{s:Intrinsic} presents and intrinsic derivation of
hypersurface deformation commutators.

\subsection{Deformation of the lapse function and shift vector from Lie derivatives}

We first derive the deformation of the lapse function and shift vector from a
coordinate transformation of space-time by using
$\delta_\epsilon g_{\mu \nu} = \mathcal{L}_\xi g_{\mu \nu}$, identifying lapse
and shift with time components of the metric according to (\ref{eq:ADM line
  element}) and using the correspondence (\ref{eq:Diffeomorphism generator
  projection}) between $\epsilon^{\mu}$ and $\xi^{\mu}$. Details of the
standard calculation of Lie derivatives can be found in Appendix~\ref{a:ADM},
in particular in equations (\ref{eq:ADM line element - Geometrodynamics}) and
(\ref{eq:ADM decomposition of Lie derivative of metric - Geometrodynamics})
which imply
\begin{eqnarray} 
    \delta_\epsilon N =
    \dot{\epsilon}^0
    + \epsilon^a \partial_a N
    - N^a \partial_a \epsilon^0
    \label{eq:Deformation of lapse - Geometrodynamics}
\end{eqnarray}
and
\begin{eqnarray}
    \delta_\epsilon N^a =
    \dot{\epsilon}^a
    + \epsilon^c \partial_c N^a
    - N^b \partial_{b} \epsilon^a
    - \sigma q^{a b} \left( \epsilon^0 \partial_b N - N \partial_b \epsilon^0 \right)
    \label{eq:Deformation of shift - Geometrodynamics}
\end{eqnarray}
 as well as
\begin{eqnarray}
    \delta_\epsilon q_{a b} =
    \frac{\epsilon^0}{N} \dot{q}_{a b}
    + \epsilon^c \partial_c q_{a b}
    + q_{c a} \partial_{b} \epsilon^c
    + q_{c b} \partial_{a} \epsilon^c
    - \frac{\epsilon^0}{N} \left( N^c \partial_c q_{a b}
    + q_{c a} \partial_{b} N^c
    + q_{c b} \partial_{a} N^c \right)
\label{eq: Deformations of spatial metric - Geometrodynamics}
\end{eqnarray}
for the spatial components of $g_{\mu\nu}$. In the first two equations, the
time derivatives of gauge functions $\epsilon^0$ and $\epsilon^a$ are defined
as
\begin{equation} \label{deltaNeps}
  \dot{\epsilon}^I={\mathcal
    L}_t\epsilon^I=\delta_{(N,N^a)}\epsilon^I
\end{equation}
using the collective index $I$ for the four components of $\epsilon$. In the
last step, the time derivative of $\epsilon^I$ obtained from (\ref{eq:ADM line
  element - Geometrodynamics}) and (\ref{eq:ADM decomposition of Lie
  derivative of metric - Geometrodynamics}) is interpreted as a specific gauge
transformation of the gauge functions. The resulting conclusion that gauge
functions are subject to gauge transformations will be justified below.

Equations~(\ref{eq:Deformation of lapse - Geometrodynamics}) and
(\ref{eq:Deformation of shift - Geometrodynamics}) match (\ref{eq:Off-shell
  gauge transformations for lapse}) and (\ref{eq:Off-shell gauge
  transformations for shift}), derived canonically from the conventional
brackets (\ref{DD})--(\ref{HH}) valid for phase-space independent lapse and
shift. The geometrical derivation, however, may be applied also to phase-space
dependent gauge functions. As general components of the space-time metric
(\ref{eq:ADM line element}), the lapse function and shift vector are
independent of the spatial metric, constituting different kinematical degrees
of freedom. However, specific space-time metrics may well use lapse or shift
depending on the spatial metric, for instance in the well-known example of a
Friedmann--Robertson--Walker metric in conformal time. The Lie-derivative
identities (\ref{eq:Deformation of lapse - Geometrodynamics}) and
(\ref{eq:Deformation of shift - Geometrodynamics}) can still be applied in
this case, provided we use the chain rule in coordinate derivatives of $N$ and
$N^a$, as for instance in
\begin{equation}
  \partial_aN=\frac{\partial N}{\partial x^a}+ \frac{\partial q_{bc}}{\partial
    x^a} \frac{\partial N}{\partial q_{bc}}
\end{equation}
if the lapse function depends on the spatial metric.  Including such extra
terms implicitly in (\ref{eq:Deformation of lapse - Geometrodynamics}) and
(\ref{eq:Deformation of shift - Geometrodynamics}) is the only difference
implied by phase-space dependent lapse and shift in the geometrical
derivation.

As usual, the lapse function and shift vector play dual roles, defining
components of the space-time metric and components of the time-evolution
vector field relative to a normal frame adapted to a foliation. The
Lie-derivative identities derived so far make use only of the former role and
require a foliation only in expressing the space-time vector field $\xi^{\mu}$
in the normal frame, implying components $\epsilon^{I}$. However, since the
same identities can be viewed as transformations of components of the
time-evolution vector field, which is a specific example of a gauge
transformation, the same equations can be rewritten as commutator
relationships between two gauge transformations, one by
$\epsilon^I_{1}=\epsilon^I$ as before and one by $\epsilon^I_2=N^I$. This
observation justifies the final equation in (\ref{deltaNeps}), subjecting
gauge functions to gauge transformations.  Simple substitution then expresses
(\ref{eq:Deformation of lapse - Geometrodynamics}) and (\ref{eq:Deformation of
  shift - Geometrodynamics}) as
the commutator
\begin{eqnarray}
    \delta_{\epsilon_1} \epsilon_2^I - \delta_{\epsilon_2} \epsilon_1^I
    = \Delta^I
    \label{eq:Deformation of generators - Geometrodynamics}
\end{eqnarray}
where
\begin{eqnarray}
    \Delta^0 
    &=&
    \epsilon^b_1 \partial_b \epsilon^0_2 - \epsilon^b_2 \partial_b \epsilon^0_1
    \label{eq:Delta generator - Geometrodynamics0}
\end{eqnarray}
and
\begin{eqnarray}
    \Delta^a
    &=&
    \epsilon^b_1 \partial_b \epsilon_2^a - \epsilon^b_2 \partial_b \epsilon_1^a
    + \sigma q^{a b} \left(\epsilon_1 \partial_b \epsilon_2^0 - \epsilon_2 \partial_b \epsilon_1^0\right)
    \ .
    \label{eq:Delta generator - Geometrodynamicsa}
\end{eqnarray}
The second term in the commutator, $-\delta_{\epsilon_2} \epsilon_1^I$, is
obtained from the $\dot{\epsilon}^I$ in (\ref{eq:Deformation of lapse -
  Geometrodynamics}) and (\ref{eq:Deformation of shift - Geometrodynamics})
after replacing $(N,N^a)$ in (\ref{deltaNeps}) with general gauge functions,
$\epsilon_2^I$.

In this way, time derivatives in (\ref{eq:Deformation of lapse -
  Geometrodynamics}) and (\ref{eq:Deformation of shift - Geometrodynamics}),
implied by the original space-time Lie derivatives, can be given an algebraic
interpretation through the commutator (\ref{eq:Deformation of generators -
  Geometrodynamics}) without using equations of motion.  Unlike the remaining
derivative (\ref{eq: Deformations of spatial metric - Geometrodynamics}) of
the spatial metric, which contains a time derivative of $q_{ab}$ rather than
of gauge functions, the transformation of lapse and shift therefore has an
off-shell correspondence with gauge transformations adapted to a foliation, or
with hypersurface deformations.  The derivations in Section~\ref{s:Intrinsic}
will relate this outcome to intrinsic properties of hypersurface deformations
by rederiving the components (\ref{eq:Delta generator - Geometrodynamics0})
and (\ref{eq:Delta generator - Geometrodynamicsa}) of the off-shell commutator
without using space-time Lie derivatives.

\subsection{Lie commutator from deformations of the gauge functions}
\label{sec:Deformation of the generators from geometrodynamics}

In our next step, we compute the Lie commutator of two space-time vector
fields, using their normal decomposition adjusted to a family of
hypersurface. To this end, we choose two space-time vector fields,
$\xi_1^\mu = \xi_1^t t^\mu + \xi_1^a d^\mu_a = \epsilon_1^0 n^\mu +
\epsilon_1^a s^\mu_a$ and
$\xi_2^\mu = \xi_2^t t^\mu + \xi_2^a d^\mu_a = \epsilon_2^0 n^\mu +
\epsilon_2^a s^\mu_a$ with components $\xi_{i}^{\mu}$ in a space-time frame
$(t^{\mu},s_a^{\mu})$ as well as corresponding gauge functions
$\epsilon_{i}^{I}$ in the normal frame.

We first apply a deformation according to $\xi_1^{\mu}$.  The components
of $\xi_2^\mu$ then receive a deformation given by the Lie derivative
$\mathcal{L}_{\xi_1} \xi_2^\mu = \xi_1^\nu \partial_\nu
\xi_2^\mu - \xi_2^\nu \partial_\nu \xi_1^\mu$. In an ADM
decomposition, using the relationship (\ref{eq:Diffeomorphism generator
  projection}) between components in a coordinate and a normal frame, we have
the components
\begin{eqnarray}
    \mathcal{L}_{\xi_1} \xi_2^t &=& 
    \xi_1^t \partial_t \xi_2^t
    - \xi_2^t \partial_t \xi_1^t
    + \xi_1^a \partial_a \xi_2^t
    - \xi_2^a \partial_a \xi_1^t
    \nonumber\\
    &=& 
    \frac{\epsilon_1^0}{N} \partial_t \left(\frac{\epsilon_2^0}{N}\right)
    - \frac{\epsilon_2^0}{N} \partial_t \left(\frac{\epsilon_1^0}{N}\right)\nonumber\\
&&    + \left( \epsilon_1^a - \frac{\epsilon_1^0}{N} N^a \right) \partial_a \left(\frac{\epsilon_2^0}{N}\right)
    - \left( \epsilon_2^a - \frac{\epsilon_2^0}{N} N^a \right) \partial_a \left(\frac{\epsilon_1^0}{N}\right)
    \nonumber\\
    &&+ \frac{1}{N} \left( \epsilon_1^a \partial_a \epsilon_2^0
    - \epsilon_2^a \partial_a \epsilon_1^0 \right)\nonumber\\
    &=&
    \frac{1}{N^2} \left( \epsilon_1^0 \left( \dot{\epsilon}_2^0
    + \epsilon_2^a \partial_a N
    - N^a \partial_a \epsilon_2^0 \right)
    - \epsilon_2^0 \left( \dot{\epsilon}_1^0
    + \epsilon_1^a \partial_a N
    - N^a \partial_a \epsilon_1^0 \right)
    \right)
    \nonumber\\
    &&+ \frac{1}{N} \left( \epsilon_1^a \partial_a \epsilon_2^0
    - \epsilon_2^a \partial_a \epsilon_1^0 \right)
    \nonumber\\
    &=&
    \frac{1}{N^2} \left( \epsilon_1^0 \delta_{\epsilon_{2}} N
    - \epsilon_2^0 \delta_{\epsilon_{1}} N
    \right)
    + \frac{1}{N} \left( \epsilon_1^a \partial_a \epsilon_2^0
    - \epsilon_2^a \partial_a \epsilon_1^0 \right)
\end{eqnarray}
using (\ref{eq:Deformation of lapse - Geometrodynamics}) in the last line.
With analogous initial steps, we also arrive at
\begin{eqnarray}
    \mathcal{L}_{\xi_1} \xi_2^a
    &=&
    \frac{1}{N} \left(
    \epsilon_1^0 \left( \dot{\epsilon}_2^a
    + \epsilon_2^b \partial_b N^a
    - N^b \partial_b \epsilon_2^a \right)
    - \epsilon_2^0 \left( \dot{\epsilon}_1^a
    + \epsilon_1^b \partial_b N^a
    - N^b \partial_b \epsilon_1^a \right) \right)
    \nonumber\\
    &&
    + \epsilon_1^b \partial_b \epsilon_2^a
    - \epsilon_2^b \partial_b \epsilon_1^a
    - N^a \mathcal{L}_{\xi_1} \xi_2^t
    \nonumber\\
    &=&
    \frac{1}{N} \left(
    \epsilon_1^0 \left( \delta_{\epsilon_2} N^a + \sigma q^{a b}
        \left(\epsilon_2^0 \partial_b N - N \partial_b \epsilon_2^0
        \right) \right)\right.\nonumber\\
  &&\left.\qquad
    - \epsilon_2^0 \left( \delta_{\epsilon_1} N^a + \sigma q^{a b} \left(\epsilon_1^0 \partial_b N - N \partial_b \epsilon_1^0 \right) \right) \right)
    \nonumber\\
    &&
    + \epsilon_1^b \partial_b \epsilon_2^a
    - \epsilon_2^b \partial_b \epsilon_1^a
    - N^a \mathcal{L}_{\xi_1} \xi_2^t
    \,.\label{eq:Lie derivative of vector - Geometrodynamics}
\end{eqnarray}
Hence,
\begin{eqnarray}
    \mathcal{L}_{\xi_1} \xi_2^\mu
    &=&
    \left(\mathcal{L}_{\xi_1} \xi_2^t\right) t^\mu
    + \left(\mathcal{L}_{\xi_1} \xi_2^a\right) s_a^\mu
    \nonumber\\
    &=&
    \left( \epsilon_1^a \partial_a \epsilon_2^0
        - \epsilon_2^a \partial_a \epsilon_1^0 \right) n^\mu\nonumber\\
  &&
    + \left(
    \epsilon_1^b \partial_b \epsilon_2^a
    - \epsilon_2^b \partial_b \epsilon_1^a
    - \sigma q^{a b} \left( \epsilon_1^0 \partial_b \epsilon_2^0
    - \epsilon_2^0 \partial_b \epsilon_1^0 \right)
    \right) s_a^\mu
    \nonumber\\
    &&
    + \epsilon_2^0 \delta_{\epsilon_1} n^\mu
    - \epsilon_1^0 \delta_{\epsilon_2} n^\mu\,.
\end{eqnarray}
In the last line, we have introduced the expression
\begin{equation} \label{deltanmu}
  \delta_{\epsilon}n^{\mu}=-\frac{\delta_{\epsilon}N}{N} n^{\mu}
  -\frac{\delta_{\epsilon}N^a}{N} s_a^{\mu}
\end{equation}
which in the next subsection will be shown to equal the gauge deformation of the
unit normal. (See equation~(\ref{eq:Deformation of normal vector - Geometrodynamics}).)
The remaining terms in the Lie derivative contain only gauge deformations of
the gauge functions and can be written as 
\begin{equation}
     \mathcal{L}_{\xi_1} \xi_2^\mu=
    \Delta^0 n^\mu
    + \Delta^a s_a^\mu
    + \epsilon_2^0 \delta_{\epsilon_1} n^\mu
    - \epsilon_1^0 \delta_{\epsilon_2} n^\mu
    \label{eq:ADM decomposition of Lie derivative of vector}
\end{equation}
with the commutator components (\ref{eq:Delta generator -
  Geometrodynamics0}) and (\ref{eq:Delta generator - Geometrodynamicsa}),
originally defined according to (\ref{eq:Deformation of generators -
  Geometrodynamics}).  The Lie derivative, initially computed in a coordinate
frame, is therefore equivalent to the gauge commutator
$ \delta_{\epsilon_1} \xi_2^\mu - \delta_{\epsilon_2} \xi_1^\mu$ computed in
the normal frame where
$\xi_{(i)}^\mu = \epsilon_{(i)}^0 n^\mu + \epsilon_{(i)}^a s_a^\mu$:
\begin{eqnarray}\label{eq:Transf of gauge functions - relation Lie and gauge commutator}
    &&\delta_{\epsilon_1} \xi_2^\mu - \delta_{\epsilon_2} \xi_1^\mu\nonumber\\
    &=&
    \left(\delta_{\epsilon_1} \epsilon^0_2
    - \delta_{\epsilon_2} \epsilon^0_1\right) n^\mu
    + \left(\delta_{\epsilon_1} \epsilon^a_2
    - \delta_{\epsilon_2} \epsilon^a_1\right) s_a^\mu
    + \epsilon^0_2 \delta_{\epsilon_1} n^\mu
    - \epsilon^0_1 \delta_{\epsilon_2} n^\mu
\end{eqnarray}
equals $\mathcal{L}_{\xi_1} \xi_2^\mu$ upon using (\ref{eq:Deformation
  of generators - Geometrodynamics}). (The Lie derivative of a vector field is
antisymmetric, such that we do not need to compute a commutator in this case.)

\subsection{Intrinsic view of hypersurface deformations}
\label{s:Intrinsic}

A family of hypersurfaces in a 4-dimensional geometry (which may have
Lorentzian or Euclidean signature) directly implies the existence of
characteristic ingredients, given by the induced geometry and extrinsic
curvature or, equivalently, the unit normal
vector field. Using only these ingredients, a deformation of an embedded
hypersurface is parameterized by components relative to the normal vector
field and a spatial basis. If one considers the commutator of two hypersurface
deformations, these ingredients in general change after the first deformation,
and the second deformation is defined relative to the transformed
ingredients. Here, we present a detailed derivation of all relevant terms,
allowing for phase-space dependent gauge functions. We begin with a discussion
of active diffeomorphism in the context of a space-time foliated into hypersurfaces.

\subsubsection{Active diffeomorphisms and Lie derivatives}

\begin{center}
\begin{figure}
  \includegraphics[width=16cm]{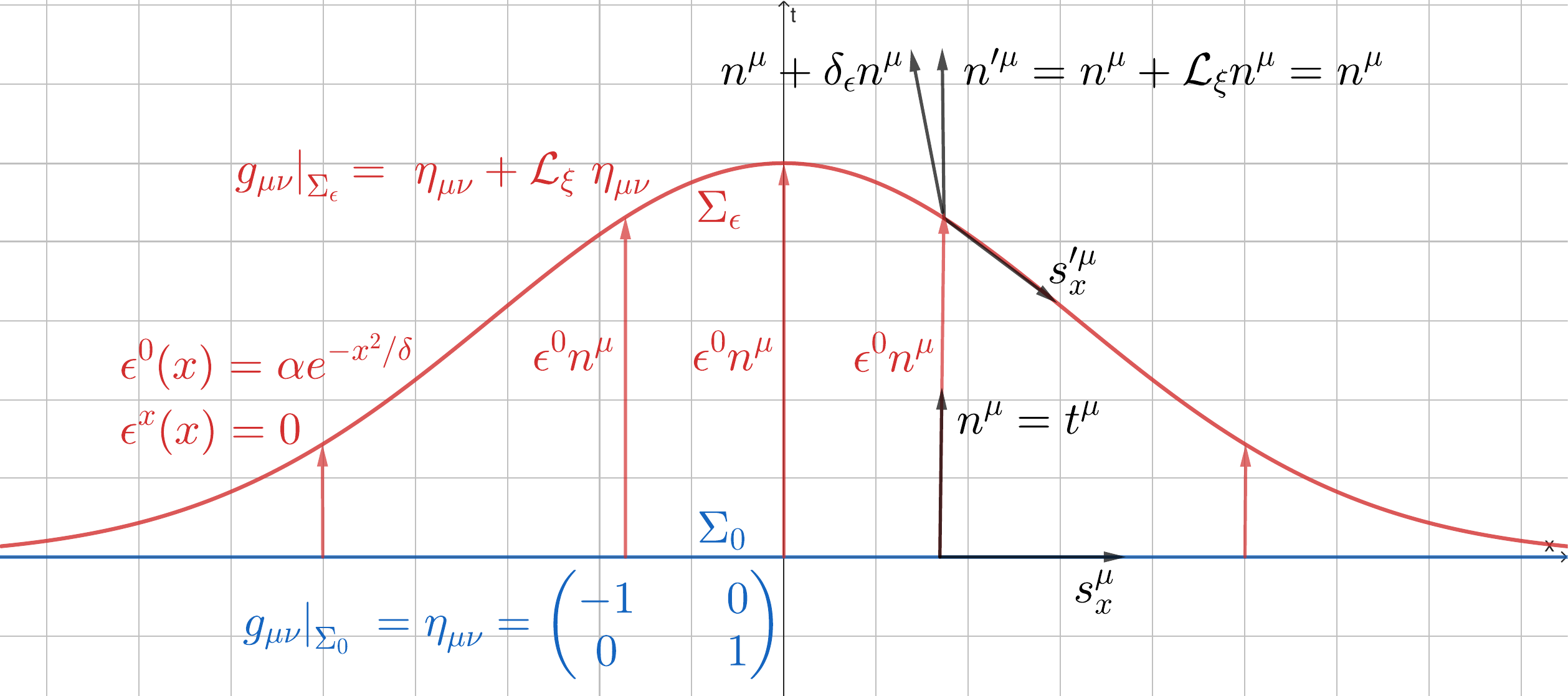}
  \caption{Deformation of a hypersurface with Lie-dragged fields. This
    transformation can be interpreted as passive because it changes the
    spatial basis within the hypersurface in order for it to remain
    tangential. The vector fields $\xi^{\mu}$ and $\epsilon^{\mu}$ are related
    by (\ref{xieps}) for $N=1$ and $N^a=0$, and therefore have identical
    components. While the induced metric and $s_x^{\mu}$ are Lie dragged by
    this transformation, $n^{\mu}$ has a vanishing Lie derivative in this
    special case, and therefore is not Lie dragged into the new directions
    required for the deformed hypersurface. The correct normal,
    $n^{\mu}+\delta_{\epsilon}^{\mu}$ can be derived as a gauge
    transformation, as shown in the main text.  (Normal directions are drawn
    assuming Lorentzian signature.) \label{fig:Passive}}
\end{figure}
\end{center}

\begin{center}
\begin{figure}
  \includegraphics[width=16cm]{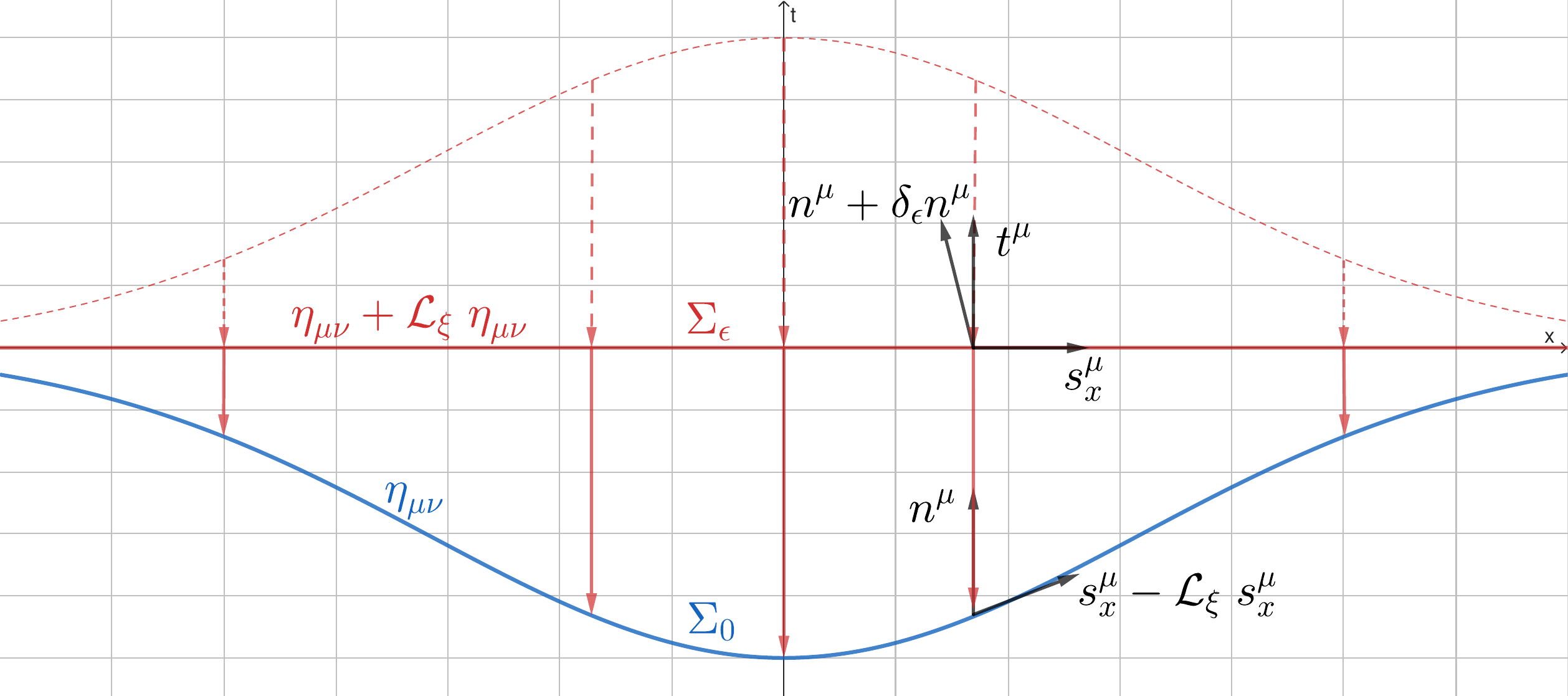}
  \caption{An active transformation that preserves the 3-dimensional manifold
    on which fields are defined. The infinitesimally deformed hypersurface is
    transformed back to the original position while dragging along fields
    defined on it. In particular, the normal vector is changed by an active
    transformation while the spatial basis is unchanged. This interpretation
    of transformations is suitable for comparisons with a canonical theory in
    which gauge flows are defined on a fixed 3-dimensional manifold.  (Normal directions
    are drawn assuming Lorentzian signature.) \label{fig:Active}}
\end{figure}
\end{center}

Active diffeomorphisms in space-time (pushforwards) drag the fields around but
leave the basis unchanged, while passive diffeomorphisms only change reference
directions in a basis. This distinction is standard for space-time theories,
but it has to be spelled out in more detail here when we are interested in
transformations of fields on hypersurfaces in space-time. We will take an
active viewpoint. However, for a comparison with canonical theories, which use
and transform fields within a fixed spatial manifold, we have to make sure
that all fields on hypersurfaces are related to the same 3-dimensional
manifold also in the geometrical description. In particular, an active
transformation should change the fields, but it should not actively deform the
hypersurface. Figures~\ref{fig:Passive} and \ref{fig:Active} illustrate this
definition of an active transformation suitable for a canonical theory.

In our notation, the space-time basis is given by $t^\mu$ and $s^\mu_a$.
Therefore, a vector field $\tau^\mu$ with components $\tau^t = 1$,
$\tau^a=(0,0,0)$ in one frame is not equivalent to the basis vector $t^\mu$
when it is transformed by an active diffeomorphism generated by $\xi^\mu$: An
active transformation does not change the basis vector $t^\mu$, but the vector
field $\tau^\mu$ experiences a transformation by
\begin{eqnarray}
    \mathcal{L}_\xi \tau^\mu = \dot{\xi}^\mu
\end{eqnarray}
which does not vanish in general.
We must therefore be careful when looking at the transformation of certain
objects.  In particular, we are interested in the transformation of the normal
vector under active diffeomorphisms.  A priori, it is not obvious whether its nature is
that of a vector field, of a basis vector, or a mix of both.  The latter case
would imply that the normal is an object that does not transform via simple
Lie derivatives, which is possible as known from the example of
connections. In what follows, we will use geometrical features of
hypersurfaces for systematic derivations of transformation properties.

\subsubsection{Deformation of the normal vector}

Intrinsic properties of hypersurface deformations depend on transformations of
the spatial metric and the normal vector field. The former are defined
directly as gauge transformations generated by the constraints, while the
latter have to be derived from conditions on the normal. We derive them in
this subsection and then use the results for an intrinsic discussion of gauge
functions and their commutator, (\ref{eq:Deformation of generators - Geometrodynamics}).

If the normal vector has a status of a space-time vector field, then under active
diffeomorphisms its transformation would be given by a Lie derivative
$n^\mu \to n^\mu + \mathcal{L}_\xi n^\mu$ with
\begin{eqnarray}
    \mathcal{L}_\xi n^\mu &=& - \frac{1}{N} \left( \dot{\epsilon}^0
    + \partial_b N
    - N^b \partial_b \epsilon^0 \right) n^\mu
    - \frac{1}{N} \left( \dot{\epsilon}^a
    + \epsilon^b \partial_b N^a
    - N^b \partial_b \epsilon^a \right) s_a^\mu
    \nonumber\\
    &=&
    - \frac{\delta_\epsilon N}{N} n^\mu
    - \left(\frac{\delta_\epsilon N^a}{N} - \frac{q^{a b} \left(\epsilon^0 \partial_b N - N \partial_b \epsilon^0 \right)}{N}\right) s_a^\mu
    \ .
\end{eqnarray}
However, recall that the basis vectors
${s_{(t_1)}}^\mu_a={s_{(t)}}^\mu_a = s^\mu_a$ are unchanged by active
diffeomorphisms in our interpretation.  If $n^\mu+\mathcal{L}_\xi n^\mu$ were
to play the role of the normal vector on the new hypersurface, then it should
be orthogonal to $s^\mu_a$ in the new hypersurface too, but this is not the
case:
\begin{eqnarray}
    g^{(t1)}_{\mu \nu} (n^\mu +\mathcal{L}_\xi n^\mu) s^\mu_b
    &=& n^\mu \mathcal{L}_\xi g^{(t)}_{\mu a} + g^{(t)}_{\mu b} \mathcal{L}_\xi n^\mu
    + O (\xi^2)
    \nonumber\\
    &=&
    \frac{q^{a c} \left(\epsilon^0 \partial_c N - N \partial_c \epsilon^0 \right)}{N}
    \ .
\end{eqnarray}
Therefore, $n^\mu+\mathcal{L}_\xi n^\mu$ cannot be the normal vector on the
new hypersurface; see also Figs.~\ref{fig:Passive} and \ref{fig:Active} for an illustration.

An arbitrary hypersurface is described by specifying its embedding in a
globally hyperbolic space-time (or, depending on the signature, 4-dimensional
space) by a mapping $y^a\to x^{\mu}$ from hypersurface coordinates $y^a$ to
space-time coordinates $x^{\mu}$. Such a hypersurface embedded in a Riemannian
manifold $M$ with metric $g_{\mu\nu}$ with signature $\sigma$ then has the
unit normal vector
\begin{equation}
    n^{\mu} =
    \frac{\sigma}{||\cdot||} g^{\mu \alpha} \epsilon_{\alpha \beta \gamma
      \delta} \frac{\partial x^\beta}{\partial y^a} \frac{\partial
      x^\gamma}{\partial y^b} \frac{\partial x^\delta}{\partial y^c} 
    \frac{\epsilon^{a b c}}{3!}=
         \frac{\sigma}{||\cdot||} \bar{n}^\mu
    \ ,
    \label{eq:Unit normal vector - general - Geometrodynamics}
\end{equation}
where we use $||\cdot||$ in order to indicate a factor required for
normalization of the remaining terms, $\bar{n}^{\mu}$, that provide the normal
direction. The factor of $\sigma$ implements the condition that the normal be
future-pointing in the Lorentzian case, $\sigma=-1$.

In a foliation, each hypersurface $\Sigma_t$ can be associated to a fixed
value of one of the coordinates, $t$, if the embedding coordinate system
$x_{(t)}\colon \Sigma_t\to M$ is adapted to the foliation such that
$x^\mu_{(t)}= (t , y^1 , y^2, y^3)$. The expression for the normal vector in
such an adapted system simplifies to
\begin{equation}
    n^{\mu}_{(t)} =
    \frac{\sigma}{||\cdot||_{(t)}} g^{\mu \alpha}_{(t)} \epsilon_{\alpha \beta
      \gamma \delta} \frac{\partial x_{(t)}^\beta}{\partial y^a}
    \frac{\partial x_{(t)}^\gamma}{\partial y^b} \frac{\partial
      x_{(t)}^\delta}{\partial y^c} 
    \frac{\epsilon^{a b c}}{3!}
= \frac{\sigma}{\sqrt{|g^{t t}_{{\rm (t)}}|}} g^{\mu t}_{(t)}
    \label{eq:Unit normal vector of foliation hypersurfaces - Geometrodynamics}
\end{equation}
with the normalization factor
$||\cdot||_{(t)} = \sqrt{|g^{t t}_{(t)}|} = 1 / N$.  Here, and in the
following, the subscript $(t)$ indicates the restriction of a function or tensor to the
hypersurface $\Sigma_t$.

To obtain the strict normal vector on the new hypersurface $n_{(t1)}^\mu$, we
use its geometric definition 
\begin{eqnarray}
    n^{\mu}_{(t1)} &=& \frac{\sigma}{\sqrt{|g^{t t}_{{\rm (t1)}}|}} g^{\mu t}_{(t1)} 
    = \sigma N \left( g^{\mu t}_{(t)} \left(1+\frac{\delta_\epsilon N}{N}\right) + \mathcal{L}_\xi g^{\mu t}_{{\rm (t)}} \right)
    + O(\epsilon^2)
\end{eqnarray}
with components
\begin{equation}
    n^t_{(t1)}= \frac{1}{N} -\frac{\delta_\epsilon N}{N^2}
    + O(\epsilon^2)
\end{equation}
\begin{equation}
    n^a_{(t1)}= - \frac{N^a}{N} n^t_{(t1)}
    - \frac{1}{N} \delta_\epsilon N^a
    + O(\epsilon^2)
    \ .
\end{equation}
They can be combined and written to the present order as
\begin{eqnarray}
    n^\mu_{(t1)} &=& n^t_{(t1)} t^\mu
    + \left(- \frac{N^a}{N} n^t_{(t1)}
    - \frac{1}{N} \delta_\epsilon N^a\right) s^\mu_a
    \nonumber\\
    &=&
    n^t_{(t1)} N n^\mu_{(t)}
    - \frac{1}{N} \delta_\epsilon N^a s^\mu_a
    \nonumber\\
    &=&
    \left(1 -\frac{\delta_\epsilon N}{N}\right) n^\mu_{(t)}
    - \frac{1}{N} \delta_\epsilon N^a s^\mu_a
    =: n^\mu_{(t)} + \delta_\epsilon n^\mu_{(t)}
\end{eqnarray}
from which
\begin{equation}
    \delta_\epsilon n^\mu_{(t)} = -\frac{\delta_\epsilon N}{N} n^\mu_{(t)}
    - \frac{\delta_\epsilon N^a}{N} s^\mu_a\,,
    \label{eq:Deformation of normal vector - Geometrodynamics}
  \end{equation}
  already referred to in (\ref{deltanmu}), directly follows.
The ratios
\begin{equation} \label{deltannormal}
  \delta_{\epsilon}^{\rm normal}n^{\mu}=-\frac{\delta_{\epsilon}N}{N}
\end{equation}
and
\begin{equation} \label{deltantangential}
  \delta_{\epsilon}^{\rm tangential}n^{\mu}=-\frac{\delta_{\epsilon}N^a}{N}
\end{equation}
are therefore the normal and tangential components of
$\delta_{\epsilon}n^{\mu}$ in the original normal frame, expressed in terms of
the temporal and spatial components, $N^{-1}$ and $-N^{-1}N^a$, in a
coordinate frame. The lapse function and shift vector inevitably appear in the
transformations of normal components because the unit normal property depends
on the space-time metric.

\subsubsection{Commutator of hypersurface deformations}

Consider two consecutive deformations with generators $\epsilon_1$ and
$\epsilon_2$.  The resulting coordinate transformation is of the form
\begin{eqnarray}
    \Delta_{1 2} x_{(t)}^\mu
    &=&
    x_{(t)}^\mu
    + \epsilon_1^0 (x_{(t)}) n_{(t)}^\mu
    + \epsilon_1^b (x_{(t)}) s^\mu_b
    + \epsilon_2^0 (x_{(t 1)}) n_{(t 1)}^\mu
    + \epsilon_2^b (x_{(t 1)}) {s_{(t 1)}}^\mu_b
    \nonumber\\
    &=&
    x_{(t)}^\mu
    + \left(\epsilon_1^0 (x_{(t)})
    + \epsilon_2^0 (x_{(t)})
    + \delta_{\epsilon_1} \epsilon_2^0 (x_{(t)})\right) n_{(t)}^\mu
    + \epsilon_2^0 (x_{(t)}) \delta_{\epsilon_1} n_{(t)}^\mu
    \nonumber\\
    &&
    + \left(\epsilon_1^b (x_{(t)})
    + \epsilon_2^b (x_{(t)})
    + \delta_{\epsilon_1} \epsilon_2^b (x_{(t)}) \right) {s_{(t)}}^\mu_b
    \ . \label{Delta}
\end{eqnarray}
If we switch the order of deformations we obtain $\Delta_{2 1} x^\mu_{(t)}$ by
simply swapping the labels 1 and 2.  The difference of the two expressions is
\begin{eqnarray}
    \Delta_{[1,2]} x^\mu_{(t)}
    &\equiv&
    \Delta_{1 2} x^\mu_{(t)}
    - \Delta_{2 1} x^\mu_{(t)}
    \nonumber\\
    &=&
    \left(\delta_{\epsilon_1} \epsilon_2^0 (x_{(t)})
    - \delta_{\epsilon_2} \epsilon_1^0 (x_{(t)}) \right) n_{(t)}^\mu
    + \left(\delta_{\epsilon_1} \epsilon_2^b (x_{(t)})
    - \delta_{\epsilon_2} \epsilon_1^b (x_{(t)}) \right) {s_{(t)}}^\mu_b
    \nonumber\\
    &&
    + \epsilon_2^0 (x_{(t)}) \delta_{\epsilon_1} n_{(t)}^\mu
    - \epsilon_1^0 (x_{(t)}) \delta_{\epsilon_2} n_{(t)}^\mu
    \nonumber\\
    &=&
    \Delta^0 n_{(t)}^\mu
    + \Delta^a {s_{(t)}}^\mu_a
    + \epsilon_2^0 (x_{(t)}) \delta_{\epsilon_1} n_{(t)}^\mu
    - \epsilon_1^0 (x_{(t)}) \delta_{\epsilon_2} n_{(t)}^\mu
    \nonumber\\
    &=&
    \mathcal{L}_{\xi_1} \xi_2^\mu (t)
    \label{eq:Commutator of hypersurface deformations - coordinates}
\end{eqnarray}
equals the Lie commutator (\ref{eq:ADM decomposition of Lie derivative of
  vector}).  Using antisymmetry, the meaning $\mathcal{L}_{\xi_1} \xi_2^\mu$
is not just the deformation of $\xi_2^\mu$ by $\xi_1^\mu$, but also the
difference of their mutual deformations. We conclude that the Lie commutator
of two space-time vector fields can be derived off-shell from hypersurface
deformations. The constraint brackets for hypersurface-deformation generators
can be recognized in the coefficients $\Delta^0$ and $\Delta^a$. However, the
Lie commutator also contains terms related to the gauge deformation of the
unit normal, which does not have a direct analog in the standard brackets used
for hypersurface deformations. 

\subsubsection{Constraint brackets from hypersurface deformations}
\label{s:brackets}

The result (\ref{eq:Commutator of hypersurface deformations - coordinates})
contains extra terms compared with the conventional canonical constraint brackets
(\ref{DD})--(\ref{HH}), given by
deformations of the normal vector. Using our transformation law for the normal
vector, these contributions can be written in terms of transformations of the
lapse function and shift vector: Continuing with (\ref{Delta}), we have
\begin{eqnarray}
    \Delta_{1 2} x_{(t)}^\mu
    &=&  x_{(t)}^\mu
    + \left(\epsilon_1^0 (x_{(t)})
    + \epsilon_2^0 (x_{(t)})
    + \delta_{\epsilon_1} \epsilon_2^0 (x_{(t)})\right) n_{(t)}^\mu
    + \epsilon_2^0 (x_{(t)}) \delta_{\epsilon_1} n_{(t)}^\mu
    \nonumber\\
    &&
    + \left(\epsilon_1^b (x_{(t)})
    + \epsilon_2^b (x_{(t)})
    + \delta_{\epsilon_1} \epsilon_2^b (x_{(t)}) \right) {s_{(t)}}^\mu_b
    \nonumber\\
    &=&
    x_{(t)}^\mu
    + \left(\epsilon_1^0 (x_{(t)})
    + \epsilon_2^0 (x_{(t)})
    + \delta_{\epsilon_1} \epsilon_2^0 (x_{(t)})
    - \epsilon_2^0 (x_{(t)}) \frac{\delta_{\epsilon_1} N}{N}
    \right) n_{(t)}^\mu
    \nonumber\\
    &&
    + \left(\epsilon_1^b (x_{(t)})
    + \epsilon_2^b (x_{(t)})
    + \delta_{\epsilon_1} \epsilon_2^b (x_{(t)})
    - \epsilon_2^0 (x_{(t)}) \frac{\delta_{\epsilon_1} N^b}{N}
    \right) {s_{(t)}}^\mu_b
    \ .
\end{eqnarray}
We switch the order of deformations to obtain $\Delta_{2 1} x^\mu_{(t)}$ by
simply swapping the labels and compute the commutator
\begin{eqnarray}
    \Delta_{[1,2]} x^\mu_{(t)}
    &=&
    \Delta_{1 2} x^\mu_{(t)}
    - \Delta_{2 1} x^\mu_{(t)}
    \nonumber\\
    &=&
    \left(\delta_{\epsilon_1} \epsilon_2^0 (x_{(t)})
    - \delta_{\epsilon_2} \epsilon_1^0 (x_{(t)})
    - \epsilon_2^0 (x_{(t)}) \frac{\delta_{\epsilon_1} N}{N}
    + \epsilon_1^0 (x_{(t)}) \frac{\delta_{\epsilon_2} N}{N}
    \right) n_{(t)}^\mu
    \nonumber\\
    &&
    + \left(\delta_{\epsilon_1} \epsilon_2^b (x_{(t)})
    - \delta_{\epsilon_2} \epsilon_1^b (x_{(t)})
    - \epsilon_2^0 (x_{(t)}) \frac{\delta_{\epsilon_1} N^b}{N}
    + \epsilon_1^0 (x_{(t)}) \frac{\delta_{\epsilon_2} N^b}{N}
    \right) {s_{(t)}}^\mu_b
    \ ,
    \label{eq:Commutator of hypersurface deformations - coordinates - Frame-fixed version}
\end{eqnarray}
where we have used (\ref{eq:Deformation of generators - Geometrodynamics}).

This commutator can be written in terms of abstract hypersurface-deformation
generators, ${\cal H}$ in the normal direction and $\vec{\cal H}$ in tangential
directions, and their commutators.  After the first deformation by
$\epsilon_1$, the second deformation by $\epsilon_2$ is
generated by
${\cal H}^{\delta_{\epsilon_1}} [\epsilon_2^0 + \delta_{\epsilon_1}
\epsilon_2^0 , \vec{\epsilon}_2 + \delta_{\epsilon_1}
\vec{\epsilon}_2 ]$ with $\delta_{\epsilon_1}$-terms adjusting the
second gauge functions to the new normal basis. To first order in
gauge functions, this expression equals
\begin{eqnarray}
&&    {\cal H}^{\delta_{\epsilon_1}} [\epsilon_2^0 + \delta_{\epsilon_1} \epsilon_2^0]
    + \vec{\cal H}^{\delta_{\epsilon_1}} \left[\vec{\epsilon}_2 +
   \delta_{\epsilon_1} \vec{\epsilon}_2\right]\nonumber\\ 
    &=&
    {\cal H} \left[ \epsilon_2^0
    + \delta_{\epsilon_1} \epsilon_2^0
    - \epsilon_2^0 \frac{\delta_{\epsilon_1} N}{N} \right]
    + \vec{\cal H}\left[ \vec{\epsilon}_2
    + \delta_{\epsilon_1}\vec{\epsilon}_2
    - \epsilon_2^0 \frac{\delta_{\epsilon_1} \vec{N}}{N} \right]
    \ .
\end{eqnarray}
The combination of two consecutive deformations is then generated by
\begin{eqnarray}
    {\cal H} \left[ \epsilon_1^0 + \epsilon_2^0
    + \delta_{\epsilon_1} \epsilon_2^0
    - \epsilon_2^0 \frac{\delta_{\epsilon_1} N}{N} \right]
    + \vec{\cal H}\left[ \vec{\epsilon}_1 + \vec{\epsilon}_2
    + \delta_{\epsilon_1} \vec{\epsilon}_2
    - \epsilon_2^0 \frac{\delta_{\epsilon_1} \vec{N}}{N} \right]
\end{eqnarray}
and their commutator has the abstract generator
\begin{eqnarray}
    &&
    {\cal H} \left[ \delta_{\epsilon_1} \epsilon_2^0
    - \delta_{\epsilon_2} \epsilon_1^0
    + \epsilon_1^0 \frac{\delta_{\epsilon_2} N}{N}
    - \epsilon_2^0 \frac{\delta_{\epsilon_1} N}{N} \right]   + \vec{\cal
       H}\left[ \delta_{\epsilon_1} \vec{\epsilon}_2 
    - \delta_{\epsilon_2} \vec{\epsilon}_1
    + \epsilon_1^0 \frac{\delta_{\epsilon_2}\vec{N}}{N}
    - \epsilon_2^0 \frac{\delta_{\epsilon_1} \vec{N}}{N} \right]
    \nonumber\\
    &=&
    {\cal H} \left[ \Delta^0
    + \frac{1}{N} \left( \epsilon_1^0 \delta_{\epsilon_2} N
    - \epsilon_2^0 \delta_{\epsilon_1} N \right) \right]
    + \vec{\cal H}\left[ \vec{\Delta}
    + \frac{1}{N} \left( \epsilon_1^0 \delta_{\epsilon_2}\vec{N}
    - \epsilon_2^0 \delta_{\epsilon_1} \vec{N}\right) \right]
\end{eqnarray}
where we used (\ref{eq:Deformation of generators - Geometrodynamics}).
This result is in agreement with the geometric expression (\ref{eq:Commutator
  of hypersurface deformations - coordinates - Frame-fixed version}) for
coordinate transformations. According to (\ref{eq:Deformation of normal vector
  - Geometrodynamics}), the commutator is expressed by the generator
\begin{equation} \label{Hepsnormal}
    {\cal H} \left[ \Delta^0
    + \epsilon_2^0 \delta_{\epsilon_1}^{\rm normal}n^{\mu}
    - \epsilon_1^0 \delta_{\epsilon_2}^{\rm normal}n^{\mu} \right]
    + \vec{\cal H}\left[ \vec{\Delta}
    + \epsilon_2^0 \delta_{\epsilon_1}^{\rm tangential}n^{\mu}
    - \epsilon_1^0 \delta_{\epsilon_2}^{\rm tangential}n^{\mu} \right]
\end{equation}
in terms of components of $\delta_{\epsilon}n^{\mu}$ in the normal basis.

The canonical hypersurface-deformation brackets are recovered by separating the
transformation of the lapse function and shift vector from that of the gauge
functions
\begin{eqnarray} \label{Hepsnormal-separation}
    {\cal H} \left[ \Delta^0
    + \epsilon_2^0 \delta_{\epsilon_1}^{\rm normal}n^{\mu}
    - \epsilon_1^0 \delta_{\epsilon_2}^{\rm normal}n^{\mu} \right] &=& H [\Delta^0] + {\cal H}_n \left[ \epsilon_2^0 \delta_{\epsilon_1}^{\rm normal}n^{\mu} - \epsilon_1^0 \delta_{\epsilon_2}^{\rm normal}n^{\mu} \right]
    \\
    \vec{\cal H}\left[ \vec{\Delta}
    + \epsilon_2^0 \delta_{\epsilon_1}^{\rm tangential}n^{\mu}
    - \epsilon_1^0 \delta_{\epsilon_2}^{\rm tangential}n^{\mu} \right] &=& \vec{H}[\vec{\Delta}] + \vec{\cal H}_n\left[ \epsilon_2^0 \delta_{\epsilon_1}^{\rm tangential}n^{\mu} - \epsilon_1^0 \delta_{\epsilon_2}^{\rm tangential}n^{\mu} \right]
    .\nonumber
\end{eqnarray}
The commutator of two spatial deformations is realized by choosing
$\epsilon_1^0 = \epsilon_2^0 = 0$:
\begin{equation}
    \left[\vec{H}[\vec{\epsilon}_1] , \vec{H}[\vec{\epsilon}_2]\right] =
    \vec{H} \left[\epsilon_1^b \partial_b \vec{\epsilon}_2 -
      \epsilon_2^b \partial_b \vec{\epsilon}_1\right] 
    \ .
    \label{eq:Spatial-spatial hypersurface deformations - Geometrodynamics}
\end{equation}
The commutator of one normal and one spatial deformation is realized by
choosing $\epsilon_1^a = 0$ and $\epsilon_2^0 = 0$:
\begin{equation}
    \left[H[\epsilon_1^0] , \vec{H}[\vec{\epsilon}_2]\right] =
    H\left[- \epsilon_2^b \partial_{b} \epsilon_1^0\right]
    \ .
    \label{eq:Normal-spatial hypersurface deformations - Geometrodynamics}
\end{equation}
And the commutator of two normal deformations is realized by choosing
$\epsilon_1^a = \epsilon_2^a = 0$: 
\begin{eqnarray}
    \left[ H[\epsilon_1^0] , H[\epsilon_2^0] \right] &=&
    H_a \left[ \sigma q^{a b} \left( \epsilon_1^0 \partial_b \epsilon_2^0 - \epsilon_2^0 \partial_b \epsilon_1^0 \right)\right]
    \ .
    \label{eq:Normal-normal hypersurface deformations - Geometrodynamics}
\end{eqnarray}

These results agree with the conventional canonical constraint brackets
(\ref{DD})--(\ref{HH}).  The normal deformations in (\ref{Hepsnormal}) do not
directly appear in the canonical brackets (\ref{DD})--(\ref{HH}) because the
former relate to the commutator
$\delta_{\epsilon_1} \xi_2^\mu - \delta_{\epsilon_2} \xi_1^\mu$ while the
latter are implied by the commutator
$\delta_{\epsilon_1} \epsilon_2^I - \delta_{\epsilon_2}
\epsilon_1^I$. These two expressions differ by the appearance of the normal
vector, as seen in (\ref{eq:Transf of gauge functions - relation Lie and gauge
  commutator}), whose components in a coordinate frame are expressed here in
terms of $N$ and $N^a$. The normal deformations in (\ref{Hepsnormal}) depend
on these components through (\ref{deltannormal}) and
(\ref{deltantangential}). Alternatively, these terms can be interpreted as an
additional dependence on the space-time metric used to implement the unit
normal property.

The relations (\ref{eq:Spatial-spatial hypersurface deformations -
  Geometrodynamics})--(\ref{eq:Normal-normal hypersurface deformations -
  Geometrodynamics}) do not depend on lapse and shift. However, in a fully
constrained canonical gauge theory, a specific choice of the gauge functions
$\epsilon^I$ may be made such that some of them are interpreted as the lapse
function and shift vector of the evolution generator.  This role distinguishes
lapse and shift among all the gauge functions.  When discussing how the normal
generator is deformed in the context of time evolution, one must take into
account the transformation of lapse and shift, just as  we
previously derived (\ref{HNeps}) and added corresponding contributions to
(\ref{Hee}). Including lapse and shift in this way completes the geometric derivation
of hypersurface deformations based on (\ref{Hepsnormal}).

\section{Canonical hypersurface deformations off-shell}
\label{s:Offshell}

As a result of the standard canonical formulation of general relativity, the
theory is generally covariant because its gauge transformations are on-shell identical
with space-time Lie derivatives of components of the space-time metric. For
this result, the theory must be taken on-shell, not only imposing the
constraints but also using equations of motion to identify which phase-space
functions can be considered time derivatives of others. However, as a gauge
theory, the canonical formulation by itself should have a suitable
interpretation of its gauge content even if the field equations are not
imposed. The classic analysis of \cite{Regained,LagrangianRegained} provides a
partial answer to this question by bringing in the important concept of
hypersurface deformations, but this original analysis had not been done fully
off-shell. Moreover, it did not provide a complete analysis of the phase-space
dependence of lapse and shift required by the presence of a structure function
in the constraint brackets.

The preceding section demonstrated crucial off-shell properties of
hypersurface deformations related to the Lie commutator of space-time vector
fields. Now, we aim to fill in the missing ingredients in traditional
discussion of canonical gravity and provide new conclusions relevant for an
interpretation of canonical quantum gravity and modified gravity in canonical
form. We will also compare the canonical structure of hypersurface
deformations with their general geometrical properties obtained in the
preceding section.

\subsection{Gauge picture of the constraint brackets}
\label{sec:Geometrodynamic meaning of the hypersurface deformation algebra}

We begin our new analysis with a set of transformation equations for all
ingredients provided by the structure of hypersurface deformations. As a new
contribution, this includes not only tensor components such as the space-time
metric but also properties of the foliation, primarily the unit normal, which,
as seen in our geometrical derivation, should change in a specific way if the
hypersurface is deformed in the active picture used here. In the off-shell
case, in which there is no relationship with Lie derivatives of the metric in
an embedding space-time geometry, formal changes of a foliation within a
4-dimensional manifold are still meaningful and can be related to gauge
transformations.

\subsubsection{Ingredients}

We will have the following ingredients at our disposal: Since we have a
canonical theory of constraints as functions on a phase space, we can make use
of the basic phase-space degrees of freedom $q_{ab}$ and $p^{ab}$ as spatial
tensors (with a density weight in the case of $p^{ab}$). In general, it is not
guaranteed that $q_{ab}$ can be interpreted as the spatial metric on a
hypersurface in an embedding space-time, and $p^{ab}$ be related to extrinsic
curvature of this hypersurface. In fact, as basic phase-space degrees of
freedom, these ingredients are defined only up to canonical
transformations. However, using constraints with brackets of the form
(\ref{DD})--(\ref{HH}), the structure function can be used for an unambiguous
definition of the spatial metric that makes it possible to interpret the
constraints as hypersurface-deformation generators. For now, analyzing the
classical theory, we could assume that the spatial metric identified in this
way is indeed one of the basic phase-space degrees of freedom, $q_{ab}$, but
this will not be necessary in the detailed derivations that follow. (This
assumption is dropped completely in emergent modified gravity
\cite{Higher,HigherCov}, to which our results therefore apply too.)
Off-shell,  the momenta $p^{ab}$ are not related to extrinsic
curvature even classically because we cannot use equations of motion to
identify time derivatives.

The canonical fields given by $q_{ab}$ and its momenta have obvious gauge
transformations generated by Poisson brackets with the constraints, but they
provide only partial information about the foliation. In the classical case,
we also need the concept of a unit normal $n^{\mu}$ such that we can define a
candidate for a space-time metric as
$g^{\mu\nu}=q^{\mu\nu}+\sigma n^{\mu}n^{\nu}$. Off-shell, in the absence of
4-dimensional Lie derivatives related to gauge transformations, we cannot
interpret $g^{\mu\nu}$ as a space-time tensor or use it to define a line
element, but it remains an important ingredient toward a well-defined
space-time geometry on-shell. This classical argument becomes circular if the
existence of a pre-defined space-time metric in which hypersurfaces are
embedded is removed: We need a space-time metric in order to define what
``unit normal'' means, and we need a unit normal in order to define a
candidate space-time metric starting with the spatial metric. Here, the
constraints help us to provide unambiguous meaning to the constructions by
reverting the classical result that the Hamiltonian constraint $H[N]$
generates deformations of a hypersurface by a multiple $N$ of the unit normal
at any given point. In the absence of an embedding geometry off-shell, we
instead define the unit normal direction as the result of a deformation
generated by $H[1]$.

This definition depends on what we consider the Hamiltonian constraint to be,
as a specific element in our constrained system. Among all the constraints, the
Hamiltonian constraint is distinguished by having a Poisson bracket (\ref{HH})
with itself in which the structure function appears. We therefore cannot
redefine the Hamiltonian constraint by arbitrary linear combinations
(although some combinations may be allowed that maintain the structure; see
Section~\ref{s:Linear}). The form of the brackets is preserved if we, for
instance, multiply the Hamiltonian constraint by a constant $\Omega$, but then
the structure function is multiplied by $\Omega^2$, and so is the inverse
spatial metric it determines. Nevertheless, if we follow our definition and
interpret a multiplication of the Hamiltonian constraint by $\Omega$ as a
rescaling of the unit normal vector to $\Omega n^{\mu}$, the space-time
geometry implied by the inverse metric
$\Omega^2q^{\mu\nu}+\sigma (\Omega n^{\mu})(\Omega
n^{\nu})=\Omega^2g^{\mu\nu}$ is unchanged because a constant $\Omega^2$ can
simply be absorbed in a coordinate transformation
$x^{\mu}\mapsto \Omega x^{\mu}$.

In the absence of an embedding geometry off-shell we therefore retain a
well-defined concept of the unit normal, based on the structure of constraint
brackets.  In general, the unit normal should be subject to change if the
hypersurface is deformed. The normal components are not phase-space functions
on which standard gauge transformations are defined by Poisson brackets with
the constraints, and therefore their changes must be derived in a different
way.  As we will show, the unit normal is nevertheless subject to unambiguous
gauge transformations that follow from the condition that it be transformed to
the new unit normal by a hypersurface deformation, such that
$n^{\mu}q_{\mu\nu}=0$, accompanied by a standard canonical gauge
transformation of the spatial tensor $q_{\mu\nu}$ induced on the
hypersurface. All the components that determine the candidate space-time
metric therefore have well-defined gauge transformations. We will first derive
this complete set of gauge transformations relevant for a space-time slicing,
and then see how it is related to brackets of the gauge generators.

\subsubsection{Canonical transformation of the normal vector}

The independent components of a candidate space-time metric are given by
$q^{ab}$ and $n^{\mu}$.  A general hypersurface deformation is labeled
by a 4-component vector $\epsilon^{\mu}$ where $\epsilon^0$ indicates the
shift in the normal direction and $\epsilon^a$, $a=1,2,3$, specifies a spatial
shift tangential to the spacelike hypersurface relative to a spatial basis
$s_a^{\mu}$ of tangent vectors. All four components may depend on coordinates
chosen on the spacelike hypersurface as well as the coordinate $t$ that labels
different slices in space-time. However, $\epsilon^{\mu}$ is not necessarily a
space-time 4-vector because its components depend on the slicing. Under such a
gauge transformation, both the unit normal vector and the spatial tensor $q_{ab}$
generically change to $n^\mu + \delta_\epsilon n^\mu$ and
$q_{a b} + \delta_{\epsilon} q_{a b}$, and one could contemplate a change
$\delta_{\epsilon}s_a^{\mu}$ of the spatial basis as well. However, we will
assume that $\delta_{\epsilon}s_a^{\mu}=0$ in order to consider only proper
deformations of hypersurfaces that do not contain a change of the spatial
basis. (A general hypersurface transformation can be written as a composition
of a proper hypersurface deformation as just defined and a purely spatial
deformation. Since purely spatial deformations have a well-understood
correspondence with infinitesimal spatial diffeomorphisms, we will not
include them here.)

In general, it is not guaranteed that the unit normal vector $n^{\mu}$ will
remain unit normal if we try to keep it constant throughout,  assuming a gauge
transformation with $\delta_{\epsilon}n^{\mu}=0$.  Since we define the
property of $n^{\mu}$ as a unit normal prior to the deformation, we have
$n^\mu q_{\mu \nu} = 0$ where $q_{\mu \nu} = q_{a b} s^a_\mu s^b_\nu$. The
property of being normal can be maintained if and only if
$\delta_\epsilon (n^\mu q_{\mu \nu}) = 0$, but since
$\delta_{\epsilon}q_{\mu\nu}$, computed directly from Poisson brackets with
the constraints, is non-zero and not necessarily normal to $n^{\mu}$, there
has to be a cancellation between $\delta_{\epsilon}q_{\mu\nu}$ and
$\delta_{\epsilon}n^{\mu}$. 

The unit normal has a strict relationship with the time-evolution vector field
$t^{\mu}$ and the spatial basis, with coefficients depending on the lapse
function and shift vector used to define the frame from which evolution is
observed. The time-evolution vector field, just like the observer's frame, is
independent of the space-time foliation and therefore not subject to gauge
transformations, and we assume that the spatial basis $s_a^{\mu}$ does not
change either. For lapse and shift, we have already obtained gauge
transformations (\ref{eq:Off-shell gauge transformations for lapse}) and
(\ref{eq:Off-shell gauge transformations for shift}) from the constraint
brackets. Using (\ref{eq:Time-evolution vector field}), it then follows
immediately that
\begin{equation}
    \delta_\epsilon n^\mu =
    - \frac{1}{N} n^{\mu}\delta_\epsilon N 
    - \frac{1}{N} s_a^{\mu}\delta_\epsilon N^a 
    \label{eq:Unit normal vector gauge transformation}
  \end{equation}
  in agreement with the geometrical result (\ref{eq:Deformation of normal
    vector - Geometrodynamics}). 
The unit normal
is therefore subject to a specific, non-vanishing gauge transformation under a
change of the foliation.

This result, together with the normal condition
$\delta_\epsilon(n^\mu q_{\mu \nu}) = 0$, implies that the gauge
transformation of $q_{\mu\nu}$ is not orthogonal to the unit normal
vector,
\begin{equation}
    N n^\nu \delta_\epsilon q_{\mu \nu} = -Nq_{\mu\nu}\delta_{\epsilon}n^{\nu}=
    q_{\mu b} \delta_\epsilon N^b \not=0
\end{equation}
in general.  The same expression can be obtained in an alternative way,
directly using the normal condition: Starting with $n^\mu q_{\mu \nu} = 0$ and
rewriting the normal vector using (\ref{eq:Time-evolution vector field}), we
have $q_{t \nu} = N^a q_{a \nu}$ (and $q_{t t} =N^aq_{at}= N^a N^b q_{a b}$),
which in turn implies
\begin{equation}
  N n^\nu \delta_\epsilon q_{\mu \nu} = \delta_\epsilon q_{\mu t} - N^b
  \delta_\epsilon q_{\mu b} = q_{\mu b} \delta_\epsilon N^b
\end{equation}
where we used the gauge-independence of $t^{\mu}$ and $s_a^{\mu}$ (for proper
hypersurface deformations) in the first step.

\subsubsection{Transformation of the normal generator}

Since the normal direction in general changes if we apply a hypersurface
deformation, the same should be the case for its generator, given as a linear combination of the
constraints. Transforming a gauge generator constitutes a passive deformation,
by which the spacetime fields (including lapse and shift as components of the
space-time metric) are gauge transformed with the opposite sign compared to
the gauge functions that determine the time-evolution generator $H[N,\vec{N}]$
as a linear combination of the constraints.

If we first perform a deformation generated by some $\epsilon_1^\mu$, the
deformation of the time-evolution
generator changes $H[N,\vec{N}]$ to
\begin{eqnarray}
    H [N + \delta_{\epsilon_1} N , \vec{N} + \delta_{\epsilon_1} \vec{N}]
     &=&
    H [N + \delta_{\epsilon_1} N ]
    + \vec{H}[ \delta_{\epsilon_1} \vec{N} ]
    + \vec{H} [\vec{N}]
    \nonumber\\
    &=:& H^{\bar{\epsilon}_1} [N]
    + \vec{H} [\vec{N}]
\end{eqnarray}
where we define the new normal generator
\begin{eqnarray} \label{HNeps}
    H^{\bar{\epsilon}_1}_{\bar{N}} &=&
     \left( 1 + \frac{\delta_{\epsilon_1} N}{N} \right)H
    +  \frac{\delta_{\epsilon_1} N^a}{N}H_a
\end{eqnarray}
in local (unsmeared) form, where bars on the left indicate 4-component
vectors. The deformation coefficients in $H^{\bar{\epsilon}_1}_{\bar{N}}-H$
equal the components of the gauge transformation (\ref{eq:Unit normal vector
  gauge transformation}) of $n^{\mu}$, up to a sign because the gauge
transformation passively changes the normal direction.

The spatial generator remains undeformed according to our assumption that we
perform a proper hypersurface deformation.  A proper deformation to a new
hypersurface therefore implies a transformation of the Hamiltonian constraint
to a specific linear combination with the diffeomorphism constraint. Vice
versa, if we can manage to replace the Hamiltonian constraint by a linear
combination with the diffeomorphism constraint such that the
hypersurface-deformation brackets retain their characteristic form, we can
define a new space-time geometry in which the normal direction has a new
meaning (on an undeformed hypersurface) compared with the geometry defined by
the original Hamiltonian constraint. This observation constitutes one of the
fundamental features of emergent modified gravity \cite{Higher,HigherCov}; see also
Section~\ref{s:Linear}.

\subsubsection{Transformation of the gauge functions}

A deformed generator of the form $H^{\bar{\epsilon}_1}_{\bar{N}}$ can be used
if we are interested in the commutator of two gauge transformations, one by
$\epsilon_1^{I}$ and one by $\epsilon_2^{I}$. If we first perform a gauge
transformation by $\epsilon_1^{I}$, the normal direction experienced by
$\epsilon_2^{I}$ changes (relative to the original normal frame). This
change can be implemented by applying $\epsilon_2^{I}$ with the generator
\begin{equation} \label{Hee}
  H^{\bar{\epsilon}_1}_{\bar{\epsilon_2}} =  \left( 1 +
    \frac{\delta_{\epsilon_1} \epsilon_2^0}{\epsilon_2^0} \right)H 
  +  \frac{\delta_{\epsilon_1} \epsilon_2^a}{\epsilon_2^0}H_a\,.
\end{equation}
(The first transformation by $\bar{\epsilon}_1$ is applied to the components
of the second transformation by $\bar{\epsilon}_2$ because it passively
changes the normal direction to which $\bar{\epsilon}_2$ refers.)  If we apply these
transformations to a phase-space function $\mathcal{O}$, the first
one implies that the phase-space function deforms into
\begin{eqnarray}
    \mathcal{O}^{(1)} &=& \mathcal{O} + \delta_{H[\bar{\epsilon}_1]} \mathcal{O}
\end{eqnarray}
where we denote the full generator in the gauge change
by $\delta_{H[\bar{\epsilon}_1]}=\delta_{H[\epsilon_1^0]+\vec{H}[\vec{\epsilon}_1]}$
because it will be different in successive transformations.  The second
deformation, relative to the new normal direction, results in
\begin{eqnarray}
    \mathcal{O}^{(1,2)}
    &=&
    \mathcal{O}^{(1)}
    + \delta_{H^{\bar{\epsilon}_1}_{\bar{\epsilon}_2}[\bar{\epsilon}_2]} \mathcal{O}^{(1)} = \mathcal{O}
    + \delta_{H[\bar{\epsilon}_1]} \mathcal{O}+ \delta_{H^{\bar{\epsilon}_1}_{\bar{\epsilon}_2}[\bar{\epsilon}_2]}(\mathcal{O}
    + \delta_{H[\bar{\epsilon}_1]}\mathcal{O})
    \nonumber\\
    &=&
    \mathcal{O}
    + \delta_{H[\bar{\epsilon}_1]} \mathcal{O}
    + \delta_{H^{\bar{\epsilon}_1}_{\bar{\epsilon}_2}[\bar{\epsilon}_2]} \mathcal{O} 
        +\delta_{H[\bar{\epsilon}_2]}\delta_{H[\bar{\epsilon}_1]} \mathcal{O}
\end{eqnarray}
if we ignore higher-order contributions in $\epsilon_i^{\mu}$ in the last term. Using
the explicit expression (\ref{Hee}) for the deformed generator, we obtain
\begin{equation}
   \mathcal{O}^{(1,2)} =
\mathcal{O}
    + \delta_{H[\bar{\epsilon}_1]} \mathcal{O}
    + \delta_{H[\bar{\epsilon}_2]} \mathcal{O} 
        +\delta_{H[\bar{\epsilon}_2]}\delta_{H[\bar{\epsilon}_1]} \mathcal{O}
    + \delta_{H [\delta_{\bar{\epsilon}_1} \bar{\epsilon}_2]}\mathcal{O}
    \,.
\end{equation}
Reversing the order of the deformations gives $\mathcal{O}^{(2,1)}$ with the
same result but the labels $1$ and $2$ flipped.  The commutator of the two
operations then equals
\begin{eqnarray}
    \mathcal{O}^{[1,2]}
    &=&
    \mathcal{O}^{(1,2)} - \mathcal{O}^{(2,1)}
    \nonumber\\
    &=& \delta_{H[\bar{\epsilon}_2]}\delta_{H[\bar{\epsilon}_1]}\mathcal{O}-
        \delta_{H[\bar{\epsilon}_1]}\delta_{H[\bar{\epsilon}_2]}\mathcal{O} -
        \delta_{H[\delta_{\bar{\epsilon}_2}\bar{\epsilon}_1-\delta_{\bar{\epsilon}_1}\bar{\epsilon}_2]}\mathcal{O}\\
    &=& -\delta_{[H[\bar{\epsilon}_1],H[\bar{\epsilon}_2]]} \mathcal{O}+
        [\mathcal{O},\delta_{H[\bar{\epsilon}_1]},\delta_{H[\bar{\epsilon}_2]}] 
 -
        \delta_{H[\delta_{\bar{\epsilon}_2}\bar{\epsilon}_1-\delta_{\bar{\epsilon}_1}\bar{\epsilon}_2]}\mathcal{O} \label{deltadelta} 
\end{eqnarray}
where we defined
\begin{equation} \label{Jac}
  [\mathcal{O},\delta_{H[\bar{\epsilon}_1]},\delta_{H[\bar{\epsilon}_2]}] =
  \delta_{H[\bar{\epsilon}_2]} \delta_{H[\bar{\epsilon}_1]} \mathcal{O}
  -\delta_{H[\bar{\epsilon}_1]} \delta_{H[\bar{\epsilon}_2]} \mathcal{O} +
  \delta_{[H[\bar{\epsilon}_1],H[\bar{\epsilon}_2]]} \mathcal{O}\,.
\end{equation}

If $\delta$ is a Lie-algebra representation of the commutator
$[H[\bar{\epsilon}_1],H[\bar{\epsilon}_2]]$, we have
$[\mathcal{O},\delta_{H[\bar{\epsilon}_1]},\delta_{H[\bar{\epsilon}_2]}]=0$. 
If gauge transformations are given by Poisson brackets, (\ref{Jac}) is turned
into the Jacobiator
\begin{equation} \label{JacPoisson}
  [\mathcal{O},\delta_{H[\bar{\epsilon}_1]},\delta_{H[\bar{\epsilon}_2]}]=
  \{\{\mathcal{O},H[\bar{\epsilon}_1]\},H[\bar{\epsilon}_2]\} -
  \{\{\mathcal{O},H[\bar{\epsilon}_2]\},H[\bar{\epsilon}_1]\} +
  \{\mathcal{O},\{H[\bar{\epsilon}_1],H[\bar{\epsilon}_2]\}\} 
\end{equation}
which always vanishes. However, if $\bar{\epsilon}_1$ and $\bar{\epsilon}_2$
are phase-space dependent, the Poisson Jacobiator also includes terms with
Poisson brackets of these functions rather than just the constraints. For a
non-zero Jacobiator in (\ref{deltadelta}), the gauge transformations should
then also include terms with derivatives of the gauge functions. In a later
section we will show that this is indeed the case.

The remaining terms in (\ref{deltadelta}) have an interesting interpretation
in the case of a vanishing Jacobiator if we write
$\delta_{\bar{\epsilon}_j}\bar{\epsilon}_i=\{\bar{\epsilon}_i,H[\bar{\epsilon}_j]\}$. We can then
combine these terms as
\begin{equation}\label{eq:Commutation of two deformations - Canonical}
  -\delta_{[H[\bar{\epsilon}_1],H[\bar{\epsilon}_2]]} \mathcal{O}
 -
        \delta_{H[\delta_{\bar{\epsilon}_2}\bar{\epsilon}_1-\delta_{\bar{\epsilon}_1}\bar{\epsilon}_2]}\mathcal{O}
        = -
      \{\mathcal{O},\{H[\bar{\epsilon}_1],H[\bar{\epsilon}_2]\}\}
      - \{\mathcal{O},H[\delta_{\bar{\epsilon}_2}\bar{\epsilon}_1-\delta_{\bar{\epsilon}_1}\bar{\epsilon}_2]\} \,.
\end{equation}
If $\bar{\epsilon}_1$ and $\bar{\epsilon}_2$ are phase-space dependent, the
Poisson bracket $\{H[\bar{\epsilon}_1],H[\bar{\epsilon}_2]\}$ in the first
term contains terms of the form (\ref{DD})--(\ref{HH}), but also new
contributions such as $H[\{\bar{\epsilon}_1,H[\bar{\epsilon}_2]\}]$ or
$H[H[\{\bar{\epsilon}_1,\bar{\epsilon}_2\}]]$. Identifying
$\{\bar{\epsilon}_1,H[\bar{\epsilon}_2]\}=
\delta_{\bar{\epsilon}_2}\bar{\epsilon}_1$, terms of the former type are
canceled by the new contribution
$H[\delta_{\bar{\epsilon}_2}\bar{\epsilon}_1-\delta_{\bar{\epsilon}_1}\bar{\epsilon}_2]$
in (\ref{eq:Commutation of two deformations - Canonical}) that results from a
transformation of the normal vector. Terms of the second type, however, remain
in the off-shell description. We will discuss phase-space dependent gauge
functions in more detail in a later part of this paper.

The constraint brackets (\ref{DD})--(\ref{HH}), as usually
provided, are valid only for phase-space independent gauge functions,
corresponding to the first term in (\ref{eq:Commutation of two deformations -
  Canonical}) which we can write as
\begin{equation}\label{deltaindep}
    \delta_{[H[\bar{\epsilon}_1],H[\bar{\epsilon}_2]]} \mathcal{O}=
    \delta_{H[\bar{\Delta}_{12}]}\mathcal{O}
\end{equation}
with
\begin{eqnarray}\label{eq:Delta generator}
    \Delta_{12}^0 
    &=&
    \epsilon^b_{1} \partial_b \epsilon^0_{2} - \epsilon^b_{2} \partial_b \epsilon^0_{1}
    \ ,
    \label{eq:Delta generator - normal}
\end{eqnarray}
and
\begin{eqnarray}
    \Delta_{12}^a
    &=&
    \epsilon^b_{1} \partial_b \epsilon_{2}^a - \epsilon^b_{2} \partial_b \epsilon_{1}^a
    + \sigma q^{a b} \left(\epsilon_{1}^0 \partial_b \epsilon_{2}^0 - \epsilon_{2}^0 \partial_b \epsilon_{1}^0\right)
    \label{eq:Delta generator - spatial}
\end{eqnarray}
read off from (\ref{DD})--(\ref{HH}).  The final step is to impose
path-independence, a condition introduced in \cite{Regained}, by setting
$\mathcal{O}^{[1,2]}=0$:
\begin{equation}
    \delta_{\epsilon_1} \epsilon_2^I - \delta_{\epsilon_2} \epsilon_1^I
    =
    \Delta_{12}^I\,,
    \label{eq:Deformation of generators - canonical}
  \end{equation}
  using (\ref{deltadelta}) and 
assuming a vanishing Jacobiator.

We note that replacing $\epsilon_1$ with $\epsilon$ and $\epsilon_2$ with $N$
and identifying $ \delta_{\bar{N}} \epsilon^I = \dot{\epsilon}^I$, we
rederive equations (\ref{eq:Off-shell gauge transformations for lapse}) and
(\ref{eq:Off-shell gauge transformations for shift}). This derivation follows
the previously mentioned argument that lapse and shift must be subject to
gauge transformations in order to ensure compatibility of evolution and gauge
changes, and it explicitly shows the role of the Jacobi identity.

\subsubsection{Jacobiator}

Structure functions in the constraint brackets contribute to the form of
iterated gauge transformations.  Using phase-space independent gauge
functions, (\ref{HD}) and (\ref{HH}) imply
\begin{eqnarray}
    &&\left\{ \left\{ H [\epsilon^0_1] , H [\epsilon^0_2] \right\} , H [\epsilon^0_3] \right\}
    =
    \left\{ \vec{H} \left[ \sigma q^{a b} \left( \epsilon^0_1 \partial_b \epsilon^0_2 - \epsilon^0_2 \partial_b \epsilon^0_1 \right)\right] , H [\epsilon^0_3] \right\}
    \nonumber\\
    &=&
    H \left[
    \sigma q^{a b} \left( \epsilon^0_1 \partial_b \epsilon^0_2 - \epsilon^0_2 \partial_b \epsilon^0_1 \right)
    \partial_a \epsilon^0_3
    \right]
    \nonumber\\
    &&
    + \int {\rm d}^3 x\ \sigma H_a (x) \left( \epsilon^0_1 (x) \partial_b \epsilon^0_2 (x) - \epsilon^0_2 (x) \partial_b \epsilon^0_1 (x) \right) \left\{ q^{a b} (x) , H [\epsilon^0_3] \right\}
    \nonumber\\
    &=&
    H \left[
    \sigma q^{a b} \left( \epsilon^0_1 \partial_b \epsilon^0_2 - \epsilon^0_2 \partial_b \epsilon^0_1 \right)
    \partial_a \epsilon^0_3
    \right]
    \nonumber\\
    &&
    + H_a \left[ \sigma \left( \epsilon^0_1 \partial_b \epsilon^0_2 - \epsilon^0_2 \partial_b \epsilon^0_1 \right) \left( Q^{a b} \epsilon^0_3
    + Q^{a b c} \partial_{c} \epsilon^0_3 + Q^{a b c d} \partial_{c}
       \partial_{d} \epsilon^0_3 +\cdots \right) \right]
\end{eqnarray}
where we defined the spatial $Q$-tensors by
\begin{equation}
  \left\{ q^{a b} (x) , H [\epsilon^0_3] \right\}= \left( Q^{a b} \epsilon^0_3
    + Q^{a b c} \partial_{c} \epsilon^0_3 + Q^{a b c d} \partial_{c}
    \partial_{d} \epsilon^0_3 +\cdots\right)\,.
\end{equation}
(For the classical Hamiltonian constraint of general relativity, only $Q^{ab}$ is non-zero because
the constraint does not depend on spatial derivatives of the momenta.)

Adding all cyclic permutations of the subscripts $1,2,3$ of $\epsilon_i^0$,
the $Q^{ab}$ and $Q^{abc}$ terms cancel out because they depend only on up to
first-order derivatives of $\epsilon_i^0$, which can be combined with the
factors in $\epsilon^0_j \partial_b \epsilon^0_k - \epsilon^0_k \partial_b
\epsilon^0_j$ from the constraint brackets. Terms with second or higher
derivatives of $\epsilon_i^0$ cannot be combined in the three cyclic
permutations, and therefore remain in the Jacobiator:
\begin{eqnarray}
    &&
    \left\{ \left\{ H [\epsilon^0_1] , H [\epsilon^0_2] \right\} , H [\epsilon^0_3] \right\}
    + \left\{ \left\{ H [\epsilon^0_2] , H [\epsilon^0_3] \right\} , H [\epsilon^0_1] \right\}
    + \left\{ \left\{ H [\epsilon^0_3] , H [\epsilon^0_1] \right\} , H [\epsilon^0_2] \right\}
    \nonumber\\
    &=&
    \sigma H_a \bigg[ Q^{a b c d} \bigg( \left( \epsilon^0_1 \partial_b \epsilon^0_2 - \epsilon^0_2 \partial_b \epsilon^0_1 \right) \partial_{c} \partial_{d} \epsilon^0_3
    + \left( \epsilon^0_2 \partial_b \epsilon^0_3 - \epsilon^0_3 \partial_b \epsilon^0_2 \right) \partial_{c} \partial_{d} \epsilon^0_1
    \nonumber\\
    &&\hspace{1cm}
    + \left( \epsilon^0_3 \partial_b \epsilon^0_1 - \epsilon^0_1 \partial_b
       \epsilon^0_3 \right) \partial_{c} \partial_{d} \epsilon^0_2 \bigg)
       +\cdots \bigg]
    \,.
    \label{eq:Off-shell Jacobiator of three H}
\end{eqnarray}
Since the Poisson bracket always obeys the Jacobi identity, we conclude that
$Q^{abcd}$ and higher versions vanish if $\delta$ is a Lie-algebra
representation of the commutator $[H[\epsilon_1],H[\epsilon_2]]$ and
hence may be imposed as a condition on the allowed $H$ generating $\delta$.
If the structure function $q^{ab}$ or a triad version is directly related to
the gravitational configuration variables, as in classical general relativity,
this condition means that the Hamiltonian constraint can only depend on the
gravitational momenta and their first spatial derivatives, but not higher
ones. If the structure function is not of this form, as in emergent modified
gravity to be discussed in more detail in Section~\ref{s:new}, implications of
the condition $Q^{abcd}=0$ depend on the specific form of $q^{ab}$ as a
phase-space function. In this case, higher derivatives of the momenta are not
necessarily excluded. If we consider abstract gauge transformations rather
than Poisson brackets, the condition $Q^{abcd}=0$ may be relaxed, allowing for
violations of the Jacobi identity off-shell. However, for abstract generators,
the relationship with higher-order derivatives of momenta is less clear.

\subsection{Phase-space dependent gauge functions}

The equations (\ref{DD})--(\ref{HH}) show that gauge functions in $H[N]$ and
$\vec{H}[\vec{N}]$ generated by iterations of the Poisson brackets depend on
the spatial metric and are therefore phase-space dependent. The dependence
obtained by iterating Poisson brackets of the constraints is of a specific
form, containing the inverse spatial metric $q^{ab}$ from (\ref{HH}) as well
as its spatial derivatives if the result of (\ref{HH}) is used in a follow-up
Poisson bracket with another $H[N]$ or $\vec{H}[\vec{N}]$. We will not
explicitly analyze this specific dependence but rather use this iterative
procedure of computing Poisson brackets to motivate an analysis of general
phase-space dependent gauge functions $N$ and $\vec{M}$ as well as
$\epsilon^0$ and $\vec{\epsilon}$. In the most general form, the dependence
may also include the momentum canonically conjugate to $q_{ab}$, or any of the
phase-space degrees of freedom in case $q_{ab}$ is not directly a
configuration variable.

The Poisson brackets of constraints with phase-space dependent gauge functions
are not of the original form (\ref{DD})--(\ref{HH}) but contain extra terms
determined by Poisson brackets of the constraints with gauge functions
\cite{BergmannKomarGroup}. Such terms may be compared with the last term in
(\ref{deltadelta}) that depends on the gauge transformation of a gauge
function, implied there by a changing normal direction after the first gauge
transformation in a commutator. We will now present and analyze the complete
off-shell brackets for phase-space dependent gauge functions, generalizing the
on-shell result of \cite{BergmannKomarGroup}.

\subsubsection{Poisson brackets with phase-space dependent gauge functions}

In order to set up our calculation, we consider two smeared expressions,
$\mathcal{O} [N] = \int {\rm d}^3 x \ \mathcal{O} (x) N(x)$ and
$\mathcal{U} [M] = \int {\rm d}^3 y \ \mathcal{U} (y) M(y)$ with phase-space
dependent smearing functions $N$ and $M$ and functions $\mathcal{O}(x)$ and
$\mathcal{U}(x)$ that depend on the phase-space variables locally in $x$. In
their Poisson bracket,
\begin{eqnarray}
&&    \{ \mathcal{O} [N] , \mathcal{U} [M] \} =
    \left\{ \int {\rm d}^3 x \ \mathcal{O} (x) N(x) , \int {\rm d}^3 y \ \mathcal{U} (y) M(y) \right\}
    \nonumber\\
    &=&
    \int {\rm d}^3 x {\rm d}^3 y \ \left\{ \mathcal{O} (x) , 
    \mathcal{U} (y) \right\} N(x) M (y)
    \nonumber\\
    &&+ \int {\rm d}^3 x {\rm d}^3 y \ \left\{ N(x) , 
    \mathcal{U} (y) \right\} \mathcal{O} (x) M (y)
    + \int {\rm d}^3 x {\rm d}^3 y \ \left\{ \mathcal{O} (x) , 
    M (y) \right\} N(x) \mathcal{U} (y)
    \nonumber\\
    &&+ \int {\rm d}^3 x {\rm d}^3 y \ \left\{ N(x) , 
    M (y) \right\} \mathcal{O} (x) \mathcal{U} (y)\,,
    \label{eq:Poisson brackets - phase space dependent contributions - Canonical}
\end{eqnarray}
only the first, conventional term,
\begin{equation}
  C [N,M] = \int {\rm d}^3 x {\rm d}^3 y \ \left\{ \mathcal{O} (x) , 
    \mathcal{U} (y) \right\} N(x) M (y)\,,
\end{equation}
appears if the smearing functions are phase-space independent.  There are
three additional contributions implied by a phase-space dependence of $N$ and
$M$, of two different types: depending on Poisson brackets of a gauge function
with a local expression, $\mathcal{O}$ or $\mathcal{U}$, and on the Poisson
bracket of two gauge functions, respectively. Depending on the derivations, it
may also be convenient to combine some of these terms, as in
\begin{eqnarray} \label{product}
  &&\int {\rm d}^3 x {\rm d}^3 y \ \left\{ \mathcal{O} (x) , 
    M (y) \right\} N(x) \mathcal{U} (y)
    + \int {\rm d}^3 x {\rm d}^3 y \ \left\{ N(x) , 
     M (y) \right\} \mathcal{O} (x) \mathcal{U} (y)\nonumber\\
  &=&  \int {\rm d}^3 y \ \left\{ \int {\rm d}^3 x\mathcal{O} (x) N(x) , 
      M (y) \right\} \mathcal{U} (y)
\end{eqnarray}
for the last two terms in (\ref{eq:Poisson brackets - phase space dependent
  contributions - Canonical}). 

As a useful example for the structure of the non-conventional terms, we assume
that the local expressions $\mathcal{O}(x)$ and $\mathcal{U}(x)$ depend only
on up to second-order derivatives of the phase-space variables. For the
purpose of this example, we call the configuration variables at a given point $Q^i$ and their
momenta $P_i$. The first non-conventional term in (\ref{eq:Poisson brackets -
  phase space dependent contributions - Canonical}) then reads
\begin{eqnarray}
    &&\int {\rm d}^3 x {\rm d}^3 y \ \left\{ N(x) , 
    \mathcal{U} (y) \right\} \mathcal{O} (x) M (y)
    \nonumber\\
    &=&
    \int {\rm d}^3 x {\rm d}^3 y {\rm d}^3 z \ \left( \frac{\delta N(x)}{\delta Q^i(z)} \frac{\delta \mathcal{U} (y)}{\delta P_i(z)}
    - \frac{\delta N(x)}{\delta P_i(z)} \frac{\delta \mathcal{U} (y)}{\delta Q^i(z)} \right) \mathcal{O} (x) M (y)
    \nonumber\\
    &=&
    \int {\rm d}^3 x \
    \left( \mathcal{O} \frac{\partial N}{\partial Q^i}
    - \partial_c \left( \mathcal{O} \frac{\partial N}{\partial (\partial_c Q^i)} \right)
    + \partial_c \partial_d \left( \mathcal{O} \frac{\partial N}{\partial
        (\partial_c \partial_d Q^i)} \right)\right)\nonumber\\
  &&\qquad\qquad\times
    \left(
    \frac{\partial \mathcal{U}}{\partial P_i} M
    - \partial_a \left( \frac{\partial \mathcal{U}}{\partial (\partial_a P_i)} M \right)
    + \partial_a \partial_b \left( \frac{\partial \mathcal{U}}{\partial (\partial_a \partial_b P_i)} M \right)
    \right)
    \nonumber\\
    &&- \int {\rm d}^3 x \ \left( \mathcal{O} \frac{\partial N}{\partial P^i}
    - \partial_c \left( \mathcal{O} \frac{\partial N}{\partial (\partial_c P^i)} \right)
    + \partial_c \partial_d \left( \mathcal{O} \frac{\partial N}{\partial (\partial_c \partial_d P^i)} \right)
       \right)\nonumber\\
  &&\qquad\qquad\left(
    \frac{\partial \mathcal{U}}{\partial Q^i} M
    - \partial_a \left( \frac{\partial \mathcal{U}}{\partial (\partial_a Q^i)} M \right)
    + \partial_a \partial_b \left( \frac{\partial \mathcal{U}}{\partial (\partial_a \partial_b Q^i)} M \right)
    \right)
    \ ,
\end{eqnarray}
and similarly for the other two lines.

\subsubsection{Complete constraint brackets}
\label{s:Complete}

The main phase-space functions of interest for hypersurface deformations are
the Hamiltonian and diffeomorphism constraints. If lapse $N$ and shift
$\vec{N}$ are fully phase-space dependent, the traditional constraint brackets
(\ref{DD})--(\ref{HH}) specify only the conventional terms.  Using
(\ref{eq:Poisson brackets - phase space dependent contributions - Canonical})
and a version of (\ref{product}), the full Poisson brackets of the
constraints, including non-conventional terms, are
\begin{eqnarray}\label{eq:Constraint algebra - phase space dependent generators - Canonical}
    \{ \vec{H} [ \vec{N}] , \vec{H} [ \vec{M} ] \}
    &=&
    - \vec{H} [\mathcal{L}_{\vec{M}} \vec{N}]\nonumber\\
    &&+ H_a [ \{ N^a ,  \vec{H}[\vec{M}] \} - \{ M^a , H_b [N^b] \}
    - H_b [\{ N^b , M^a \}] ]
    \ , \\
    \{ H [ N ] , \vec{H} [ \vec{M}]\}
    &=&
        - H [ M^b \partial_b N ] \nonumber\\
  &&    + H[ \{ N , \vec{H} [ \vec{M}]\}] - H_b [\{ M^b , H [N] \} + H [\{ N , M^b \}]]
    \ , \\
    \{ H [ N ] , H [ M ] \}
    &=&
    \sigma\vec{H} [ q^{a b} ( N \partial_b M - M \partial_b N )]\nonumber\\
   && + H \left[ \{N , H[M]\} - \{M , H[N]\} - H[\{N , M\}] \right] \label{HHGen}
    \,.
\end{eqnarray}
Partial derivatives $\partial_a$ by coordinates, acting on the gauge
functions, are implied by integrations by parts as in the conventional
derivation. If lapse and shift are phase-space dependent, they have to be
understood according to the chain rule, such that
\begin{equation}
  \partial_a = \frac{\partial}{\partial x^a}+\frac{\partial q_{bc}}{\partial
    x^a}\frac{\partial}{\partial q_{bc}}+ \frac{\partial p^{bc}}{\partial
      x^a}\frac{\partial}{\partial p^{bc}}\,.
\end{equation}
The last term in each constraint bracket is quadratic in the local constraints
and had been ignored in the on-shell discussion of
\cite{BergmannKomarGroup}. The quadratic nature implies that these
contributions may be interpreted as additional constraint terms with structure
functions that vanish on-shell but may be important off-shell.

As an immediate implication, we have to update equations
(\ref{eq:Off-shell gauge transformations for lapse}) and (\ref{eq:Off-shell
  gauge transformations for shift}) for the gauge transformations of lapse and
shift because they had been derived from the conventional constraint
brackets. Including non-conventional terms, the off-shell gauge transformations of
phase-space dependent lapse and shift are given by
\begin{eqnarray}
    \not\!\delta_\epsilon N &=&
    \frac{\partial \epsilon^0}{\partial t}
    + \epsilon^b \partial_b N
    - N^b \partial_b \epsilon^0\nonumber\\
&&    + \{\epsilon^0 , H[N]+H_b[N^b]\}
    - \{N , H[\epsilon^0]+H_b[\epsilon^b]\}
    + H[\{N,\epsilon^0\}] \label{deltaNfull}
\end{eqnarray}
and
\begin{eqnarray}
    \not\!\delta_\epsilon N^a &=&
    \frac{\partial\epsilon^a}{\partial t}
    + \epsilon^b \partial_b N^a
    - N^b \partial_b \epsilon^a
    - \sigma q^{a b} ( N \partial_b \epsilon^0 - \epsilon^0 \partial_b N )
    \nonumber\\
    &&
    + \{ \epsilon^a , H[N]+H_b[N^b] \}
    - \{ N^a , H[\epsilon^0]+H_b[\epsilon^b] \}
     \nonumber\\
    &&
   + H[\{N , \epsilon^a\} - \{\epsilon^0 , N^a\}]
    + H_b [\{ N^b , \epsilon^a \}]
    \,, \label{deltaNafull}
\end{eqnarray}
which can be read off from the constraint brackets as in (\ref{eq:Delta
  generator - normal}) and (\ref{eq:Delta generator - spatial}).
The contributions
$ \{\epsilon^0 , H[N]+H_b[N^b]\} - \{N , H[\epsilon^0]+H_b[\epsilon^b]\}$ to
(\ref{deltaNfull}) and
$\{ \epsilon^a , H[N]+H_b[N^b] \} - \{ N^a , H[\epsilon]+H_b[\epsilon^b] \}$
to (\ref{deltaNafull}) are of the form
$\delta_{\bar{N}}\bar{\epsilon}-\delta_{\bar{\epsilon}}\bar{N}$ as seen in
(\ref{deltadelta}). The additional terms $H[\{N,\epsilon^0\}]$ in
(\ref{deltaNfull}) and
$H[\{N , \epsilon^a\} - \{\epsilon^0 , N^a\}] + H_b [\{ N^b , \epsilon^a \}]$
in (\ref{deltaNafull}) appear only off-shell.

In (\ref{deltaNfull}) and (\ref{deltaNafull}) we have written $\not\!\delta$
because these terms do not constitute the full gauge transformations, denoted
by $\delta$, if lapse and shift are phase-space dependent: In general, the
components $N^I$ have a mixed nature as phase-space dependent quantities and
as gauge functions.  The gauge transformation of their phase-space dependence
is taken into account by the standard Poisson brackets
$\{N^I , H[\bar{\epsilon}]\}$, while their role as gauge functions must be
obtained indirectly by invoking compatibility of gauge transformations with
evolution. The second condition is implemented by the above procedure,
resulting in (\ref{deltaNfull}) and (\ref{deltaNafull}).

Taking both contributions into account, the full gauge transformation is given by
\begin{equation}
    \delta_\epsilon N^I = \not\!\delta_\epsilon N^I + \{N^I ,
    H[\bar{\epsilon}]\} ,\label{eq:Full gauge transformation}\,. 
\end{equation}
We can also define a complete time derivative as
\begin{equation}
    \frac{{\rm d} \epsilon^I}{{\rm d} t} = \frac{\partial\epsilon^I}{\partial t}
    + \{\epsilon^I , H[\bar{N}]\}
\end{equation}
in order to mimic the same combination for phase-space dependent gauge
functions $\epsilon^I$.  The results (\ref{deltaNfull}) and
(\ref{deltaNafull}) are then written as
\begin{eqnarray}
    \delta_\epsilon N &=&
    \frac{{\rm d} \epsilon^0}{{\rm d} t}
    + \epsilon^b \partial_b N
    - N^b \partial_b \epsilon^0
    + H[\{N,\epsilon^0\}] \label{deltaNfull-formal}
\end{eqnarray}
and
\begin{eqnarray}
    \delta_\epsilon N^a &=&
    \frac{{\rm d} \epsilon^a}{{\rm d} t}
    + \epsilon^b \partial_b N^a
    - N^b \partial_b \epsilon^a
    - \sigma q^{a b} ( N \partial_b \epsilon^0 - \epsilon^0 \partial_b N )
     \nonumber\\
    &&
   + H[\{N , \epsilon^a\} - \{\epsilon^0 , N^a\}]
    + H_b [\{ N^b , \epsilon^a \}]
    \,, \label{deltaNafull-formal}
\end{eqnarray}
formally taking the same form of (\ref{eq:Off-shell gauge transformations for lapse}) and (\ref{eq:Off-shell
  gauge transformations for shift}) up to the terms proportional to the constraints, which vanish on-shell.

We conclude that the lapse and shift transformations (\ref{deltaNfull-formal})
and (\ref{deltaNafull-formal}) derived from constraint brackets with
phase-space dependent lapse and shift are compatible with the geometric
derivation. They somewhat obscure the geometry because they combine all
possible terms without distinguishing between implications of the Lie
derivative, (\ref{eq:Deformation of lapse - Geometrodynamics}) and
(\ref{eq:Deformation of shift - Geometrodynamics}), and deformations of the
normal vector. But at least on-shell, they imply the correct contributions
even for phase-space dependent gauge functions. However, their off-shell
behavior does not appear to be fully geometrical because they contain extra
terms, $H[\{N , \epsilon^0\}]$ in the lapse transformation and
$H[\{N , \epsilon^a\} - \{\epsilon^0 , N^a\}] + H_b [\{ N^b , \epsilon^a \}]$
in the shift transformation, that are not reproduced at all in the geometrical
derivation.

\subsubsection{Transformation of phase-space dependent generators and consistency
  of gauge transformations}
\label{s:Dependent}

Our derivation of equation~(\ref{eq:Deformation of generators - canonical}) used in the preceding comparison
had not taken into account a potential phase-space dependence of the gauge
functions $\bar{\epsilon}$.  We now return to the procedure of
Section~\ref{sec:Geometrodynamic meaning of the hypersurface deformation
  algebra} but allow for an unrestricted phase-space dependence of the gauge
functions in two consecutive gauge transformations given by $\epsilon_{1}$ and
$\epsilon_{2}$, respectively.  Assuming a vanishing Jacobiator, we recall the
results (\ref{deltadelta}) and (\ref{eq:Commutation of two deformations -
  Canonical}):
\begin{eqnarray} 
    \mathcal{O}^{[1,2]}
    &=& -\delta_{[H[\bar{\epsilon}_1],H[\bar{\epsilon}_2]]} \mathcal{O}
    - \delta_{H[\not\delta_{\epsilon_2}\bar{\epsilon}_1-\not\delta_{\epsilon_1}\bar{\epsilon}_2]}\mathcal{O}
        \nonumber\\
  &=&    -
      \{\mathcal{O},\{H[\bar{\epsilon}_1],H[\bar{\epsilon}_2]\}\}-\{\mathcal{O},H[ \not\!\delta_{\epsilon_2}\bar{\epsilon}_1-\not\!\delta_{\epsilon_1}\bar{\epsilon}_2]\} 
\end{eqnarray}
for the commutator of two gauge transformations acting on $\mathcal{O}$.  As
already used in the preceding subsection, we define the formal gauge
transformation $\not\!\delta_{\epsilon}$ by ignoring the phase-space dependence
of $\epsilon$:
\begin{eqnarray}
    \delta_\epsilon \mathcal{O}
    &=&
    \int {\rm d}^3 x\ \left( \epsilon (x) \{ \mathcal{O} , H (x) \}
    + H (x) \{ \mathcal{O} , \epsilon (x) \} \right)
    =: \not{\!\delta}_\epsilon \mathcal{O}
    + \int {\rm d}^3 x\ H (x) \{ \mathcal{O} , \epsilon (x) \}
    \label{eq:Full gauge transformation definition}
\end{eqnarray}

For phase-space dependent gauge functions, the complete
hypersurface-deformation brackets turn this result into
\begin{equation}
    \mathcal{O}^{[1,2]}=-
    \{ \mathcal{O} , H_I [\Delta_{12}^I
    + \{\epsilon_1^I,H[\bar{\epsilon}_2]\}
      - \{\epsilon_2^I,H[\bar{\epsilon}_1]\}
    - H_J [\{ \epsilon_1^I , \epsilon_2^J \} ]] \}
 - \{\mathcal{O},H[\not\!\delta_{\epsilon_2}\bar{\epsilon}_1-\not\!\delta_{\epsilon_1}\bar{\epsilon}_2]\} .
    \label{eq:Deformation of generators - canonical - Phase space dependent generators}
\end{equation}
Here, we use the compact notation $\epsilon_2^JH_J= \int{\rm
  d}x(H(x)\epsilon^0(x)+H_a(x)\epsilon^a(x))$ with integration of local
functions implied by the summation convention. 
Imposing path-independence $\mathcal{O}^{[1,2]}=0$ we now have
\begin{equation}
  \delta_{[H[\bar{\epsilon}_1],H[\bar{\epsilon}_2]]}= \delta_{H_I[\Delta_{12}^I
  + \{\epsilon_1^I,H[\bar{\epsilon}_2]\}
  - \{\epsilon_2^I,H[\bar{\epsilon}_1]\}
    - H_J [\{ \epsilon_1^I , \epsilon_2^J \} ]]} =\delta_{H[\not\delta_{\epsilon_2}\bar{\epsilon}_1-\not\delta_{\epsilon_1}\bar{\epsilon}_2]}
\end{equation}
with
\begin{equation} \label{epsilonoffshell}
  \not\!\delta_{\epsilon_1}\epsilon_2^I
  - \not\!\delta_{\epsilon_2} \epsilon_1^I
  = \Delta_{12}^I
  + \{\epsilon_1^I,H[\bar{\epsilon}_2]\}
  - \{\epsilon_2^I,H[\bar{\epsilon}_1]\}
  - H_J [\{ \epsilon_1^I , \epsilon_2^J \} ] 
\end{equation}
as a generalization of (\ref{eq:Deformation of generators - canonical}) to
phase-space dependent gauge functions, containing an off-shell
contribution. Comparing again with the geometrical structure of hypersurface
deformations, we see that (\ref{epsilonoffshell}) matches (\ref{eq:Delta
  generator - Geometrodynamics0}) and (\ref{eq:Delta generator -
  Geometrodynamicsa}) as well as (\ref{eq:Transf of gauge functions - relation
  Lie and gauge commutator}) only on-shell, even though the geometrical
derivation of hypersurface deformations in a generic space-time (not subject
to any field equations) is valid also off-shell. There is therefore an
off-shell difference between the canonical and geometrical pictures.

We should also revisit the Jacobiator, defined in (\ref{Jac}), which now
translates into a Poisson Jacobiator with several terms compared with
(\ref{JacPoisson}):
\begin{eqnarray} \label{Jacepsilon}
[\mathcal{O},\delta_{H[\bar{\epsilon}_1]},\delta_{H[\bar{\epsilon}_2]}] &=&
  [\mathcal{O},\not\!\delta_{H[\bar{\epsilon}_1]},\not\!\delta_{H[\bar{\epsilon}_2]}]
  + [\mathcal{O} , \not\!\delta_{H[\bar{\epsilon}_1]} , \epsilon_2^J] H_J\nonumber\\
  &&+ [\mathcal{O}, \epsilon_1^I , \not\!\delta_{H[\bar{\epsilon}_2]}] H_I
  +H_IH_J
  \{\mathcal{O},\epsilon_1^I,\epsilon_2^J\}\,.
\end{eqnarray}
The last term in (\ref{Jacepsilon}) is a Jacobiator of the Poisson bracket and
vanishes. The two middle terms vanish on-shell, but they are relevant for the
off-shell behavior of hypersurface deformations.  If the gauge transformation
is applied to a smeared object, $\mathcal{O} = F [\bar{\epsilon}_3]$, then
\begin{eqnarray}
[\mathcal{O},\not\!\delta_{H[\bar{\epsilon}_1]},\not\!\delta_{H[\bar{\epsilon}_2]}]=
  \epsilon_3^K[F_K,\not\!\delta_{H[\bar{\epsilon}_1]},\not\!\delta_{H[\bar{\epsilon}_2]}]+
  F_K[\epsilon_3^K,\not\!\delta_{H[\bar{\epsilon}_1]},\not\!\delta_{H[\bar{\epsilon}_2]}] 
\end{eqnarray}
and we obtain a second relevant on-shell term unless $F_K$ is a linear
combination of the constraints $H_K$.

\section{Applications}

We have arrived at a detailed characterization of off-shell geometrical
hypersurface deformations with phase-space dependent gauge functions,
completing the previously available analysis based on \cite{Regained}. We
discussed how the resulting equations are reproduced by various terms in a
canonical derivation at least on-shell. We also showed that the canonical
equations have additional off-shell terms for phase-space dependent lapse and
shift that do not reproduce intrinsic features of the geometrical derivation.
These results have several far-reaching implications to new classes of
modified gravity theories as well as to constructions in canonical quantum
gravity.

\subsection{Geometrical conditions and new theories of emergent modified gravity}
\label{s:new}

Our analysis extends previous classic work such as
\cite{BergmannKomarGroup,Regained} to the general case of phase-space
dependent gauge functions, taking into account necessary changes of the unit
normal under hypersurface deformations.
We have observed not only off-shell
deviations between canonical brackets and geometrical hypersurface
deformations but also more subtle on-shell differences once lapse and shift are allowed to
be phase-space dependent. While additional terms appear in deformation
commutators in both cases, they are not identical owing to the fact that the
geometrical derivation takes into account changes of the normal vector after a
deformation.  As a consequence, components of the space-time metric (used to
define the unit normal property) appear in some terms, while the canonical
treatment can only refer to the gauge functions that label deformation
generators and the induced spatial metric on phase space.
An extended phase space, in which lapse and shift are accompanied by momenta, would give
indirect access to all space-time metric components in a canonical
treatment. It can be used to derive terms such as $\dot{\epsilon}^0$ in
(\ref{eq:Deformation of lapse - Geometrodynamics}) from Poisson brackets
\cite{LapseGauge}, but since it would simply add linear terms in the new
momenta to the constraints, it could not imply terms non-linear in the lapse
function as required for the full (\ref{eq:Commutator of hypersurface
  deformations - coordinates - Frame-fixed version}).  Moreover, as shown in
\cite{Regained}, it is possible to reconstruct space-time geometry without
using an extended phase space. There should therefore be a geometrical
interpretation of canonical theories even in the absence of a complete
agreement between canonical and geometrical commutators.

From this viewpoint, our results complete the foundations of emergent modified
gravity, which has recently shown how consistent space-time geometries can be
reconstructed from canonical theories \cite{Higher,HigherCov} without using some of
the implicit assuptions made in \cite{Regained}. As a consequence, new
gravitational theories have been identified, so far with a restriction to
spherical symmetry. Our constructions here allow us to generalize some of the
conditions for covariant theories to full gravity without symmetry assumptions.

\subsubsection{Emergent modified gravity}

The conventional constraint brackets (\ref{DD})--(\ref{HH}) imply geometrical
hypersurface deformations for phase-space independent gauge functions, forming
the central feature of space-time geometry derived from a canonical
formulation \cite{Regained}. For phase-space dependent gauge functions, the
constraint brackets receive additional terms of two different kinds, relevant
for on-shell and off-shell properties, respectively. The former had been
derived in \cite{BergmannKomarGroup} but not directly related to geometrical
hypersurface deformations. While some of these terms resemble features of
geometrical contributions to hypersurface deformations implied by a changing
normal direction, they are not exactly of the required form. The off-shell
terms, which had not been considered in \cite{BergmannKomarGroup}, also fail
to correspond to geometrical contributions. We conclude that the full
canonical constraint brackets with phase-space dependent gauge functions
implement geometrical hypersurface deformations neither off-shell nor on-shell.

A geometrical interpretation of the canonical theory therefore has to proceed
along the following lines: Given a canonical theory, classical or modified, we
first construct a candidate embedding space-time in which hypersurface
deformations are equivalent to gauge transformations generated by the
constraints with phase-space independent gauge functions and lapse and
shift. Once a consistent space-time geometry has been obtained, the solution
space can be extended to phase-space dependent lapse functions and shift
vectors as components of the space-time metric. The canonical theory therefore
allows the full freedom in making gauge choices as known from classical
general relativity, including cases such as conformal time in which the lapse
function is not independent of the spatial metric. However, this freedom is
realized only on the level of solutions in a compatible space-time
geometry. It does not extend to fundamental gauge transformations of the
canonical theory.

There is an additional caveat because constraint brackets of the form
(\ref{DD})--(\ref{HH}) do not always guarantee the existence of a compatible
space-time metric, as shown explicitly in emergent modified gravity. There is
an additional covariance condition that requires the structure function in
(\ref{HH}), derived from a modified canonical theory, to transform under gauge
transformations as an inverse spatial metric does under coordinate
transformations. The existence of a compatible space-time geometry in a
modified canonical theory therefore relies on three non-trivial conditions:
(i) The modified constraints must remain first class (or be
anomaly-free). (ii) The modified constraints with phase-space independent
gauge functions must obey brackets of the precise form (\ref{DD})--(\ref{HH})
as required for hypersurface deformations. (iii) The structure function in
(\ref{HH}) must on-shell transform under gauge transformations as required for
an inverse spatial metric. The latter condition requires going fully on-shell,
solving the constraints and implementing equations of motion. Once this
condition as been implemented, the momenta on phase space or the time
derivative of the spatial metric also transform as required, as shown in
\cite{HigherCov}.

Phase-space dependent gauge functions, or lapse and shift, do not play a role
in this reconstruction because canonical constraint brackets in this case do
not match hypersurface deformations even classically. Nevertheless,
phase-space dependent gauge functions can be used in an important way in
constructions of new modified theories of gravity. Starting with the classical
theory and the usual constraints $H[N]$ and $H_a[N^a]$, phase-space dependent
$N$ and $N^a$, under certain conditions, make it possible to construct sets of
constraints that obey (\ref{DD})--(\ref{HH}) with a modified structure
function. In particular, inserting in the classical $H[N]$ and $H_a[N^a]$
linear combinations $N=B \tilde{N}$ and $N^a=A^a\tilde{N}+\tilde{N}^a$, with
phase-space dependent $A^a$ and $B$ but phase-space independent $\tilde{N}$
and $\tilde{N}^a$, implies extra terms in the constraint brackets according to
(\ref{HHGen}). In general, the corresponding gauge transformations no longer
correspond to geometrical hypersurface deformations. However, if specific conditions on
$A^a$ and $B$ are satisfied, the full generator
$H[N]+H_a[N^a] = \bar{H}[\tilde{N}]+H_a[\tilde{H}^a]$ with
$\tilde{H}=BH+A^aH_a$ obeys hypersurface-deformation brackets as required for
phase-space independent $\tilde{N}$ and $\tilde{N}^a$, but with a modified
structure function. If an additional condition is obeyed, the modified
structure function transforms under gauge transformations of the new system as
required for an inverse spatial metric. A new space-time geometry is then
obtained, which need not be equivalent to the classical theory because
hypersurface deformations (still with their classical structure in terms of
brackets) are realized in an inequivalent way, or as a different geometrical
sector within the full canonical brackets obtained with phase-space dependent
lapse and shift.

\subsubsection{Gauge conditions}

Examples in spherical symmetry have already demonstrated the possibility of
new modified theories of gravity with a generally covariant space-time
interpretation. Our new methods developed here can be used to
generalize the required conditions, derived in \cite{HigherCov} from direct
computations of Poisson brackets with spherically symmetric Hamiltonian and
diffeomorphism constraints, to the full theory without symmetry conditions.

Starting with generic phase-space dependent gauge functions, our result
(\ref{eq:Deformation of generators - canonical - Phase space dependent
  generators}) together with (\ref{eq:Off-shell gauge transformations for
  lapse}) and (\ref{eq:Off-shell gauge transformations for shift}) implies,
using the definition (\ref{eq:Full gauge transformation definition}) of
$\not{\!\delta}_{\bar{\epsilon}}$, the geometric conditions
\begin{eqnarray}
    \not{\!\delta}_{\bar{\epsilon}} N
    + \{ N , H [\bar{\epsilon}] \}
    &=&
    \frac{\partial \epsilon^0}{\partial t}
    + \{ \epsilon^0 , H[\bar{N}] \}
    + \epsilon^a \partial_a N
    - N^a \partial_a \epsilon^0
\end{eqnarray}
and
\begin{eqnarray}
    \not{\!\delta}_{\bar{\epsilon}} N^a
    + \{ N^a , H [\bar{\epsilon}] \}
    &=&
    \frac{\partial \epsilon^a}{\partial t}
        + \{ \epsilon^a , H [\bar{N}] \}\nonumber\\
  &&
    + \epsilon^b \partial_b N^a
    - N^b \partial_b \epsilon^a
    - \sigma q^{a b} \left(\epsilon^0 \partial_b N - N \partial_b \epsilon^0 \right)
    \,,
\end{eqnarray}
which should hold on-shell.  These conditions restrict the allowed phase-space
dependence of lapse and shift. An extreme example is given by only phase-space
dependent lapse and shift without explicit coordinate dependence, in which
case
$\not{\!\delta}_{\bar{\epsilon}} N = \not{\!\delta}_{\bar{\epsilon}} N^a =
0$. If we then choose phase-space independent gauge functions
$\bar{\epsilon}$, it is impossible to recover the right-hand sides from the
left-hand sides for arbitrary $\bar{\epsilon}$.  Thus, for a fully covariant
theory (rather than specific solutions), lapse and shift cannot be pure
phase-space functions and must always retain a gauge residue that will
complete the geometric conditions.

The geometric conditions are non-linear if they are viewed as equations for
both $\bar{N}$ and $\bar{\epsilon}$.  Instead of trying to solve them from
scratch, it is easier to construct interesting examples with phase-space as
well as coordinate dependence of lapse and shift and the gauge functions by
starting with phase-space independent versions, $\tilde{\bar{N}}$ and
$\tilde{\bar{\epsilon}}$ , and multiplying them by phase-space functions.  For
instance, consider $N = B \tilde{N}$ and $\epsilon^0 = B \tilde{\epsilon}^0$
with phase space dependent $B$ and phase space independent $\tilde{N}$ and
$\tilde{\epsilon}^0$. Thus, on-shell
\begin{eqnarray}
&&    B \not{\!\delta}_{\bar{\epsilon}} \tilde{N}
    + \tilde{N} \{ B , H[B\tilde{\epsilon}^0]+H_a[\epsilon^a] \}\nonumber\\
    &=&
    B \frac{\partial \tilde{\epsilon}^0}{\partial t}
    + \tilde{\epsilon}^0 \{ B , H[B\tilde{N}]+H_a[N^a] \}
    + \epsilon^a \partial_a (B \tilde{N})
    - N^a \partial_a (B \tilde{\epsilon}^0)
    \ .
\end{eqnarray}
If $B$ is a spatial scalar, the on-shell condition reduces to
\begin{eqnarray}
    \not{\!\delta}_{\epsilon} \tilde{N}
    &=&
    \frac{\partial \tilde{\epsilon}^0}{\partial t}
    + \epsilon^a \partial_a \tilde{N}
    - N^a \partial_a \tilde{\epsilon}^0
    + \frac{\tilde{\epsilon}^0 \{ B , H [B \tilde{N}] \}
    - \tilde{N} \{ B , H [B \tilde{\epsilon}^0] \}}{B}
    \ .
\end{eqnarray}
If we expand
\begin{eqnarray}
    \{ B , H [\tilde{\epsilon}^0] \} 
    \big|_{{\rm O.S.}}
    &=&\left(
    {\cal B} \tilde{\epsilon}^0
    + {\cal B}^a \partial_a \tilde{\epsilon}^0
    \right)\big|_{{\rm O.S.}}
    \ ,
\end{eqnarray}
we obtain
\begin{eqnarray}
    \not{\!\delta}_{\epsilon} \tilde{N}
    \big|_{{\rm O.S.}}
    &=&\left.\left(
    \frac{\partial \tilde{\epsilon}^0}{\partial t}
    + \left(\epsilon^a + \tilde{\epsilon}^0 B {\cal B}^a\right) \partial_a \tilde{N}
    - \left(N^a + \tilde{N} B {\cal B}^a\right) \partial_a \tilde{\epsilon}^0
    \right)\right|_{{\rm O.S.}}
    \ .
\end{eqnarray}

\subsubsection{Covariance condition}

The canonical theory is covariant if and only if
\begin{eqnarray}
    \delta_\epsilon g_{\mu \nu} |_{{\rm O.S.}} = \mathcal{L}_{\xi} g_{\mu \nu} |_{{\rm O.S.}}
\end{eqnarray}
where $\epsilon^{\mu}$ and $\xi^{\mu}$ are related by (\ref{eq:Diffeomorphism
  generator projection}).  Performing the ADM decomposition of this covariance
condition using (\ref{eq:ADM decomposition of Lie derivative of metric -
  Geometrodynamics}) from Appendix~\ref{a:ADM}, it follows that this general
condition directly implies the transformations (\ref{eq:Off-shell gauge
  transformations for lapse}) and (\ref{eq:Off-shell gauge transformations for
  shift}) for lapse and shift, while the spatial metric must obey
\begin{eqnarray}
    \frac{1}{\epsilon^0} \{ q^{a b} , H [\epsilon^0]\} \big|_{{\rm O.S.}}
    &=&
    \frac{1}{N} \{ q^{a b} , \bar{H} [N]\} \big|_{{\rm O.S.}}
    \label{eq:Spatial covariance condition - first reduced form}
\end{eqnarray}
if we assume that the diffeomorphism constraint remains unmodified.

As a local functional of $\epsilon^0$, the
normal gauge transformation of the spatial metric takes the generic form
\begin{eqnarray}
    \{q_{a b} , H [\epsilon^0]\} &=&
    Q_{a b} \epsilon^0
    + Q_{a b}^c \partial_c \epsilon^0
    + Q_{a b}^{c d} \partial_c \partial_d \epsilon^0
    + \cdots
    \label{eq:Generic normal transformation of 3-metric}
\end{eqnarray}
where the spatial $Q$-tensors are phase-space dependent and the series truncates at
some finite order.
Substituting this expansion into (\ref{eq:Spatial covariance condition - first reduced form}), we obtain
\begin{eqnarray}
    Q_{a b}^c \frac{\partial_c \epsilon^0}{\epsilon^0}
    + Q_{a b}^{c d} \frac{\partial_c \partial_d \epsilon^0}{\epsilon^0}
    + \cdots \bigg|_{{\rm O.S.}}
    &=&
    Q_{a b}^c \frac{\partial_c N}{N}
    + Q_{a b}^{c d} \frac{\partial_c \partial_d N}{N}
    + \cdots \bigg|_{{\rm O.S.}}
    \label{eq:Spatial covariance condition - second reduced form}
\end{eqnarray}
(neglecting boundary terms that may result after integrating by parts).  For a
generally covariant theory, the gauge functions
$(\epsilon^0,\epsilon^a)$ must be independent of each other, and of $(N,N^a)$
as well as the phase-space degrees of freedom that they are supposed to transform.  Thus,
each $Q$-tensor in (\ref{eq:Generic normal transformation of 3-metric}) must
vanish independently, implying a series of conditions on the gauge
transformation of the spatial metric or its inverse:
\begin{eqnarray}
    \frac{\partial (\delta_{\epsilon^0} q^{a b})}{\partial (\partial_c
  \epsilon^0)} \bigg|_{{\rm O.S.}}
    = \frac{\partial (\delta_{\epsilon^0} q^{a b})}{\partial
  (\partial_c \partial_d \epsilon^0)} \bigg|_{{\rm O.S.}}
    = \cdots
    = 0
    \ .
    \label{eq:Covariance condition - modified - reduced}
\end{eqnarray}

If the covariance condition (\ref{eq:Covariance condition - modified -
  reduced}) is extended to hold off-shell, then it is a stronger condition
than the off-shell vanishing of the Jacobiator.  Indeed, if the covariance
condition is satisifed off-shell, then the Jacobiator vanishes as is readily
seen from (\ref{eq:Off-shell Jacobiator of three H}).  This result would
ensure that both the spatial metric\textemdash via (\ref{eq:Covariance
  condition - modified - reduced})\textemdash and the lapse function and shift
vector\textemdash via (\ref{eq:Off-shell gauge transformations for lapse}) and
(\ref{eq:Off-shell gauge transformations for shift})\textemdash have the
correct off-shell transformations, that is, their gauge transformations
correspond to standard coordinate transformations.  Thus, because
(\ref{eq:Off-shell gauge transformations for lapse}) and (\ref{eq:Off-shell
  gauge transformations for shift}) ensure that the equations of motion,
generated by gauge transformations, are gauge covariant, and because
(\ref{eq:Covariance condition - modified - reduced}) ensures that the gauge
transformations are equivalent to coordinate transformations, the equations of
motions are covariant.  The theory is then covariant according to standard
terminology. Nevertheless, the conditions derived here are more general than
usual arguments that lead to covariant action principles with fully contracted
space-time tensors because they allow for equations of motion that are not
necessarily derivable from an action principle.  Our geometric and covariance
conditions are therefore weaker than demanding the existence of an invariant
action functional, and thus lead to more general equations of motion that are
still covariant.

\subsubsection{Linear combination}
\label{s:Linear}

It is possible to evaluate the geometric and covariance conditions if we start
with the classical theory and implement phase-space dependent linear
combinations of lapse and shift such that the new gauge transformations still
correspond to hypersurface deformations.

Starting with a phase-space independent lapse function $\tilde{N}$, shift vector
$\tilde{N}^a$ and gauge functions $\tilde{\epsilon}^0$ and $\tilde{\epsilon}^a$, we
introduce new expressions $N = B \tilde{N}$, $N^a = A^a \tilde{N} + \tilde{N}^a$ and
$\epsilon^0 = B \tilde{\epsilon}^0$, and
$\epsilon^a = A^a \tilde{\epsilon}^0 + \tilde{\epsilon}^a$ with phase space
dependent functions $B\not=0$ (a spatial scalar) and $A^a$ (components of a
spatial vector). The geometric condition then implies
\begin{eqnarray}
    \not{\!\delta}_{\epsilon} \tilde{N}
    &=&
   \frac{\partial \tilde{\epsilon}^0}{\partial t}
    + \left( A^a \tilde{\epsilon}^0 + \tilde{\epsilon}^a \right) \partial_a \tilde{N}
    - \left( A^a \tilde{N} + \tilde{N}^a \right) \partial_a \tilde{\epsilon}^0\nonumber\\
&&   + \frac{\tilde{\epsilon}^0 \{ B , H [B \tilde{N}] \}
    - \tilde{N} \{ B , H [B \tilde{\epsilon}^0] \}}{B}
\end{eqnarray}
and
\begin{eqnarray}
    \not{\!\delta}_{\epsilon} \tilde{N}^a
    &=&
    \frac{\partial \tilde{\epsilon}^a}{\partial t}
    + \tilde{\epsilon}^b \partial_b \tilde{N}^a
    - \tilde{N}^b \partial_b \tilde{\epsilon}^a
    - \sigma B q^{a b} \left(\tilde{\epsilon}^0 \partial_b \tilde{N} - \tilde{N} \partial_b \tilde{\epsilon}^0 \right)
    \nonumber\\
    &&
    + \tilde{\epsilon}^0 \{ A^a , H [B \tilde{N}] \}
    - \tilde{N} \{ A^a , H [B \tilde{\epsilon}^0] \}
    \nonumber\\
    &&
    - A^a \left( A^b \tilde{\epsilon}^0 \partial_b \tilde{N}
    - A^b \tilde{N} \partial_b \tilde{\epsilon}^0
    + \frac{\tilde{\epsilon}^0 \{ B , H [B \tilde{N}] \}
    - \tilde{N} \{ B , H [B \tilde{\epsilon}^0] \}}{B}\right)
\end{eqnarray}
on-shell.
The function $\tilde{N}$ can thus transform exactly as the original $N$ (but
using $\tilde{\epsilon}$ instead of $\epsilon$) only if
\begin{eqnarray}
    0
    &=&
    A^b \tilde{\epsilon}^0 \partial_b \tilde{N}
    - A^b \tilde{N} \partial_b \tilde{\epsilon}^0
    + \frac{\tilde{\epsilon}^0 \{ B , H [B \tilde{N}] \}
    - \tilde{N} \{ B , H [B \tilde{\epsilon}^0] \}}{B}
\end{eqnarray}
on-shell, that is, only if $B$ is such that
\begin{eqnarray}
    \{ B , H [\tilde{\epsilon}^0] \} 
    \big|_{{\rm O.S.}}
    &=&
    \left({\cal B} \tilde{\epsilon}^0
    + {\cal B}^a \partial_a \tilde{\epsilon}^0
    \right)\big|_{{\rm O.S.}}
    \label{eq:Transformation of B - Geometric condition - Linear combination}
\end{eqnarray}
with all higher-derivative terms vanishing, and
\begin{eqnarray}
    A^a 
    \big|_{{\rm O.S.}} 
    &=& - {\cal B}^a 
    \big|_{{\rm O.S.}}
    \label{eq:A - Geometric condition - Linear combination}
    \ .
\end{eqnarray}
If these equations are fulfilled, $\tilde{N}^a$ transforms in the standard way
too (again, using $\tilde{\epsilon}$) provided that
\begin{eqnarray}
    \{ A^a , H [\tilde{\epsilon}^0] \}
    \big|_{{\rm O.S.}} 
    &=&
    \left({\cal A}^a \tilde{\epsilon}^0
    + {\cal A}^{a b} \partial_b \tilde{\epsilon}^0
    \right)\big|_{{\rm O.S.}} 
    \label{eq:Transformation of A - Geometric condition - Linear combination}
\end{eqnarray}
with all higher-derivative terms vanishing.

If the conditions on $A^a$ and $B$ can be extended to hold off-shell, an
evaluation of the Poisson bracket (\ref{HHGen}) of two Hamiltonian constraints
with phase-space dependent lapse and shift implies that the structure function
in the classical bracket is replaced by
\begin{eqnarray}
    \tilde{q}^{a b} &=&
    B^2 q^{a b} - \sigma B {\cal A}^{a b}
    \ ,
    \label{eq:New structure function - Linear combination}
\end{eqnarray}
imposing the additional condition ${\cal A}^{a b} = {\cal A}^{b a}$ for
$\tilde{q}^{a b}$ to be symmetric. 

If all these conditions are met, the redefinition of lapse and shift as well
as the gauge functions provides a candidate for a consistent geometric theory
of a space-time line element
\begin{eqnarray}
    {\rm d} \tilde{s}^2 &=&
    \sigma \tilde{N}^2 {\rm d} t^2 + \tilde{q}_{a b} ( {\rm d} x^a + \tilde{N}^a
                          {\rm d} t ) ( {\rm d} x^b + \tilde{N}^b {\rm d} t )
\end{eqnarray}
with constraints, equations of motion and gauge transformations generated by the expressions
\begin{eqnarray}
    \tilde{H} = B H + A^a H_a
    \ , \hspace{1cm}
    \tilde{H}_a = H_a
    \label{eq:New constraints - Linear combination}
\end{eqnarray}
smeared with $\tilde{\epsilon}^0$ and $\tilde{\epsilon}^a$. From the point of view of
hypersurface deformations, the new theory makes use of a new definition of the
normal vector $\tilde{n}^{\mu}$ through the redefined $\tilde{H}$, as well as a
new spatial metric $\tilde{q}_{ab}$ which together determine the space-time
metric $\tilde{g}^{\mu\nu}=\tilde{q}^{ab}s_a^{\mu}s_b^{\nu} +
\tilde{n}^{\mu}\tilde{n}^{\nu}$. 

Consistency of the new gauge transformations and an unambiguous derivation of
the new structure function require that the new constraints
(\ref{eq:New constraints - Linear combination}) have off-shell Poisson
brackets of the required form. Direct calculations show that this property is
guaranteed if the conditions on $A^a$ and $B$ hold not only on-shell, as
required for the geometric condition, but also off-shell. It is convenient to
break the full calculations of Poisson brackets up into smaller contributions,
as outlined in Appendix~\ref{a:Brackets}.
With these intermediate results, the Poisson bracket of two diffeomorphism
constraints is trivially unchanged, while we derive that
\begin{eqnarray}
    \{ \tilde{H} [ \tilde{N} ] , \vec{\tilde{H}} [ \vec{\tilde{M}}]\}
    &=&
    \{ H [ B \tilde{N}] + H_b [ A^b \tilde{N} ] , \tilde{H}_a [\tilde{M}^a] \}
    \nonumber\\
    &=&
    - H [ B \tilde{M}^a \partial_a \tilde{N} ]
    - H_a [ A^a \tilde{M}^b \partial_b \tilde{N} ]
    \nonumber\\
    &=&
    - \tilde{H} [ \tilde{M}^b \partial_b \tilde{N} ]
\end{eqnarray}
is also of the required form for a Hamiltonian and a diffeomorphism
constraint. We have
\begin{eqnarray}
    \{ \tilde{H} [ \tilde{N} ] , \tilde{H} [ \tilde{M} ]\}
    &=&
    \{ H [ B \tilde{N} ] + H_a [A^a \tilde{N}] , H [ B \tilde{M} ] + H_a [A^a \tilde{M}] \}
    \nonumber\\
    &=&
    \tilde{H}_a [ \sigma B^2 q^{a b} \left( \tilde{M} \partial_b \tilde{N} - \tilde{N} \partial_b \tilde{M} \right) ]
    + H [ B A^b \left( \tilde{N} \partial_b \tilde{M} - \tilde{M} \partial_b \tilde{N} \right) ]
    \nonumber\\
    &&
    + \int {\rm d}^3 x {\rm d}^3 y\ H (x) \bigg( 
    \{ B (x) , H (y) \} \left( \tilde{N} (x) B (y) \tilde{M} (y) - \tilde{M} (x) B (y) \tilde{N} (y) \right)
    \nonumber\\
    &&
    + \{ B (x) , B (y) \} \tilde{N} (x) H (y) \tilde{M} (y)
    \nonumber\\
    &&
    - \{ A^a (y) , H (x) \} B (x) \left( \tilde{N} (x) H_a (y) \tilde{M} (y) - \tilde{M} (x) H_a (y) \tilde{N} (y) \right)
    \nonumber\\
    &&
    + \{ B (x) , A^a (y) \} H (x) \left(\tilde{N} (x) H_a (y) \tilde{M} (y) - \tilde{M} (x) H_a (y) \tilde{N} (y) \right)
    \nonumber\\
    &&
    + \{ A^a (x) , A^b (y) \} H_a (x) \tilde{N} (x) H_b (y) \tilde{M} (y)
    \bigg)
\end{eqnarray}
for two Hamiltonian constraints.

For the last expression to be of hypersurface-deformation form, the $H$-term
in the second line must be canceled by some of the other terms, and all the
$\tilde{H}_a$-terms must combine to a single such term with a potentially
modified structure function $\tilde{q}^{ab}$.  It is precisely the off-shell
extension of the conditions (\ref{eq:Transformation of B - Geometric condition
  - Linear combination}), (\ref{eq:A - Geometric condition - Linear
  combination}), and (\ref{eq:Transformation of A - Geometric condition -
  Linear combination}) that fulfills the latter condition: We can then rewrite
\begin{eqnarray}
&&    \{ \tilde{H} [ \tilde{N} ] , \tilde{H} [ \tilde{M} ]\}\nonumber\\
    &=&
    \tilde{H}_a \left[ \sigma \left( B^2 q^{a b} - \sigma B \mathcal{A}^{a b} \right) \left( \tilde{M} \partial_b \tilde{N} - \tilde{N} \partial_b \tilde{M} \right) \right]
    - H \left[ B \left( A^b + \mathcal{B}^b \right) \left( \tilde{M} \partial_b \tilde{N} - \tilde{N} \partial_b \tilde{M} \right) \right]
    \nonumber\\
    &&
    + \int {\rm d}^3 x {\rm d}^3 y\ \bigg(
    H (x) \{ B (x) , B (y) \} \tilde{N} (x) H (y) \tilde{M} (y)
    \nonumber\\
    &&
    + \{ B (x) , A^a (y) \} H (x) \left(\tilde{N} (x) H_a (y) \tilde{M} (y) - \tilde{M} (x) H_a (y) \tilde{N} (y) \right)
    \nonumber\\
    &&
    + \{ A^a (x) , A^b (y) \} H_a (x) \tilde{N} (x) H_b (y) \tilde{M} (y)
    \bigg)
    \nonumber\\
    &=&
    \tilde{H}_a \left[ \sigma \tilde{q}^{a b} \left( \tilde{M} \partial_b \tilde{N} - \tilde{N} \partial_b \tilde{M} \right) \right]
    \nonumber\\
    &&
    + \int {\rm d}^3 x {\rm d}^3 y\ \bigg(
    H (x) \{ B (x) , B (y) \} \tilde{N} (x) H (y) \tilde{M} (y)
    \nonumber\\
    &&
    + \{ B (x) , A^a (y) \} H (x) \left(\tilde{N} (x) H_a (y) \tilde{M} (y) - \tilde{M} (x) H_a (y) \tilde{N} (y) \right)
    \nonumber\\
    &&
    + \{ A^a (x) , A^b (y) \} H_a (x) \tilde{N} (x) H_b (y) \tilde{M} (y)
    \bigg)
    \ .
\end{eqnarray}
with structure function (\ref{eq:New structure function - Linear
  combination}). The entire bracket is of hypersurface-deformation form
provided the last three lines vanish off-shell, imposing additional conditions
on the function $B$ and the vector $A^a$.
    
We can now explicitly check the role of the covariance condition
\begin{eqnarray}
    \frac{1}{\tilde{\epsilon}^0} \{ \tilde{q}^{a b} , \tilde{H} [\tilde{\epsilon}^0]\} \big|_{{\rm O.S.}}
    &=&
    \frac{1}{\tilde{N}} \{ \tilde{q}^{a b} , \tilde{H} [\tilde{N}]\} \big|_{{\rm O.S.}}
    \,.
\end{eqnarray}
Using the classical result
\begin{eqnarray}
    \{ q^{a b} , H[\tilde{\epsilon}^0]\}
    =: Q^{a b} \tilde{\epsilon}^0
\end{eqnarray}
without contributions from  derivatives of $\tilde{\epsilon}$ because the
classical Hamiltonian constraint does not depend on derivatives of the
momenta, we compute the left-hand side as
\begin{eqnarray}
   \{ \tilde{q}^{a b} , \tilde{H} [\tilde{\epsilon}^0]\} 
    &=&
    \{ \tilde{q}^{a b} , H [B \tilde{\epsilon}^0] + H_c [A^c \tilde{\epsilon}^0]\} 
    =
    \{ \tilde{q}^{a b} , H [B \tilde{\epsilon}^0] \}
    + \mathcal{L}_{\vec{A} \tilde{\epsilon}^0} \tilde{q}^{a b}
    \nonumber\\
    &=&
    \{ B^2 q^{a b} - \sigma B \mathcal{A}^{a b} , H [B \tilde{\epsilon}^0]\}
    + \mathcal{L}_{\vec{A} \tilde{\epsilon}^0} \tilde{q}^{a b}
    \nonumber\\
    &=&
    B^2 \{ q^{a b} , H [B \tilde{\epsilon}^0]\}
    + q^{a b} \{ B^2 , H [B \tilde{\epsilon}^0]\}\nonumber\\
&&    - \sigma B \{ \mathcal{A}^{a b} , H [B \tilde{\epsilon}^0]\}
    - \sigma \mathcal{A}^{a b} \{ B , H [B \tilde{\epsilon}^0]\}
    + \mathcal{L}_{\vec{A} \tilde{\epsilon}^0} \tilde{q}^{a b}
    \nonumber\\
    &=&
    \mathcal{L}_{\vec{A} \tilde{\epsilon}^0} \tilde{q}^{a b}
    + \int {\rm d}^3 y\ \bigg(
    B^2 \left( B (y) \tilde{\epsilon}^0 (y) \{ q^{a b} , H (y) \} \right)
    \nonumber\\
    &&
    + \left( 2 B q^{a b} - \sigma \mathcal{A}^{a b} \right) \left( B (y) \tilde{\epsilon}^0 (y) \{ B , H (y) \} \right)
    - \sigma B \left( B (y) \tilde{\epsilon}^0 (y) \{ \mathcal{A}^{a b} , H (y) \} \right)
    \bigg)
    \nonumber\\
    &=&
    \mathcal{L}_{\vec{A} \tilde{\epsilon}^0} \tilde{q}^{a b}
    + B^3 Q^{a b} \tilde{\epsilon}^0
    + \left( 2 B q^{a b} - \sigma \mathcal{A}^{a b} \right) \left( B {\cal B} \tilde{\epsilon}^0
    + {\cal B}^c \partial_c \left(B \tilde{\epsilon}^0\right)\right)
    \nonumber\\
    &&
    - \sigma B \int {\rm d}^3 y\ B (y) \tilde{\epsilon}^0 (y) \{ \mathcal{A}^{a b} , H (y) \}
\end{eqnarray}
where all equalities hold on-shell. If we define
\begin{eqnarray}
    \{ \mathcal{A}^{a b} , H [\tilde{\epsilon}^0] \} &=&
    \Lambda^{a b} \tilde{\epsilon}^0
    + \Lambda^{a b c} \partial_c \tilde{\epsilon}^0
\end{eqnarray}
where the higher-derivative terms are required to vanish, the covariance condition requires
\begin{eqnarray}
 &&   \Lambda^{a b c}\nonumber\\
    &=&
    \sigma B^{-2} \left( A^b \left(B^2 q^{a c} - \sigma B {\cal A}^{a c}\right)
    + A^a \left(B^2 q^{b c} - \sigma B {\cal A}^{b c}\right)
    + \left( 2 B q^{a b} - \sigma \mathcal{A}^{a b} \right) B {\cal B}^c \right)
    \nonumber\\
    &=&
    B^{-1} \left( {\cal B}^a \left( {\cal A}^{b c} - \sigma B q^{b c} \right)
    + {\cal B}^b \left( {\cal A}^{c a} - \sigma B q^{c a} \right)
    - {\cal B}^c \left( \mathcal{A}^{a b} - 2 \sigma B q^{a b} \right) \right)
\end{eqnarray}
on-shell.

These derivations show that all the covariance conditions derived explicitly
in \cite{HigherCov} for spherically symmetric models can be consistently
generalized to emergent modified gravity without symmetry conditions.  As a
corollary, an off-shell partial Abelianization may be obtained by evaluating
conditions such that the new structure function (\ref{eq:New structure
  function - Linear combination}) is zero:
\begin{eqnarray}
    {\cal A}^{a b}
    &=&
    \sigma B q^{a b}
    \,.
    \label{eq:Partial Abelianization condition}
\end{eqnarray}
This equation presents a single condition for the function $B$, recalling that
${\cal A}^{a b}$ is obtained from $B$ via (\ref{eq:A - Geometric condition -
  Linear combination}) and (\ref{eq:Transformation of A - Geometric condition
  - Linear combination}). The simplest non-trivial example is instructive:
Taking $A^a=0$ and a space-time scalar $B=1/\Omega$, as the conditions are satisfied
with the new structure function being
$\tilde{q}^{ab}=\Omega^{-2}q^{ab}$. Writing the modified emergent line element
\begin{eqnarray}
  {\rm d}\tilde{s}^2&=&\sigma \tilde{N}^2{\rm d}t^2+\tilde{q}_{ab}({\rm
                        d}x^a+N^a{\rm d}t)({\rm d}x^b+N^b{\rm d}t)\\
  &=&
  \Omega^2(\sigma N^2{\rm d}t^2+q_{ab}({\rm
    d}x^a+N^a{\rm d}t)({\rm d}x^b+N^b{\rm d}t))=\Omega^2{\rm d}s^2\,, \nonumber
\end{eqnarray}
we identify this procedure as a Weyl (or conformal) transformation which is in
general not equivalent to a coordinate transformation and defines a new
geometry. If no canonical transformation exists that turns the new spatial
metric into a configuration variable, for instance if $\Omega$ depends on
spatial derivatives, then it is the simplest non-trivial case
of emergent modified gravity.

\subsection{Implications for space-time structure in canonical quantum gravity}
\label{s:QG}

Conceptually, differences between canonical brackets of the constraints and
geometrical properties of hypersurface deformations are relevant for
conditions and interpretations to be imposed on candidates for canonical
quantum gravity. If one takes the canonical brackets seriously as an
infinitesimal description of some algebraic object in the spirit of the
``Bergmann--Komar group,'' one encounters serious mathematical and technical
challenges. Suitable algebraic descriptions \cite{ConsAlgebroid,ConsRinehart}
of this structure are more complicated than what is known for other gauge
theories, necessitated by the appearance of structure functions. And a formal
quantization not only of the constraints but also of phase-space dependent
(turned into operator-valued) gauge functions would require specific
commutator relationships between infinitely many constraints. Experience in
canonical quantum gravity makes it clear that the endeavor of
constructing such operators would be exceedingly challenging.

Our results show that such a treatment would be over-constraining because
phase-space dependent gauge functions, even though they may be suggested by a
formal iteration of Poisson brackets with structure functions, are in fact
irrelevant for a geometrical interpretation of the theory in terms of
hypersurface deformations. This conclusion is implied by off-shell differences
between canonical brackets and geometrical deformations of
hypersurfaces. Similarly, direct exponentiation of constraint operators, as
recently advocated in \cite{U1Averaging} together with an averaging procedure
to solve the constraints, makes implicit use of phase-space dependent gauge
functions in the terms of a Taylor series, which does not represent
geometrical hypersurface deformations according to our new results.

Geometrically, results in emergent modified gravity have shown, on a classical
level, that a canonical theory in which all phase-space dependent gauge
functions are considered is much larger than what is required for a theory of
Lorentzian-signature space-time. For instance, it is possible to construct
solutions with signature change \cite{EmergentSig}, such that some regions of
the emergent 4-dimensional geometry are Euclidean rather than Lorentzian, and
some of these theories are completely Euclidean. Moreover, the algebraic
property of a partial Abelianization with vanishing structure function,
implementing the condition (\ref{eq:Partial Abelianization condition}), can
geometrically be interpreted as an emergent 4-dimensional geometry with a
Carrollian structure \cite{Carrollian,CarrollianADM}. If one insists on
representing all constraint brackets with phase-space dependent gauge
functions on a quantum level, the resulting theory would not only describe
quantum space-time, but also 4-dimensional quantum space as well as Carrollian
geometries. Given the well-known and immense difficulties encountered in constructing
anomaly-free quantum theories of the gravitational constraints, it is unlikely
that such an encompassing theory can be found. Fortunately, it is not
necessary to do so if one is interested only in the physical 4-dimensional
geometries that have at least one extended region of Lorentzian signature.

We are led to the following proposal of a new interpretation of the tasks
required for canonical quantum gravity: Instead of pursuing the challenging
endeavor of finding consistent operator versions of all the constraint
brackets with phase-space dependent lapse and shift (trying to construct a
Bergmann--Komar group), it is sufficient to implement quantum constraints for
phase-space independent (non-operator valued) phase-space functions. While
some of the conditions that would otherwise be imposed on brackets or
commutators phase-space dependent gauge functions are relaxed, a covariant
quantum theory of gravity requires a previously unrecognized operator version
of the covariance condition (\ref{eq:Spatial covariance condition - first
  reduced form}), possibly with suitable generalizations if the diffeomorphism
constraint receives quantum corrections. (As shown in \cite{HigherCov}, it is
sufficient to impose a covariance condition on the spatial metric. In an
anomaly-free theory, the corresponding tensor transformation property then
follows for the momenta.)

If these conditions can be fulfilled, a consistent geometrical space-time
interpretation exists as an effective theory, using the methods of emergent
modified gravity. Phase-space dependent gauge functions (such as a lapse
function for conformal time) may then be used in the derivation of effective
space-time solutions, giving full access to the usual solution procedures of
classical gravity. But phase-space dependent lapse and shift, or their
operator versions, do not appear on a fundamental quantum level. Constructions
of quantum constraint operators such as
\cite{TwoPlusOneDef,TwoPlusOneDef2,AnoFreeWeak,AnoFreeWeakDiff} then simplify
(compared with full phase-space dependence) although the covariance condition
remains to be implemented. Depending on the outcome, these constructions may
have to be transformed back to hypersurface-deformation generators if, as is
often the case, quantum operators had been constructed by redefining the
original brackets with phase-space independent gauge functions, for instance
to eliminate structure functions. 

\section{Discussion}

Our analysis of canonical and geometrical versions of hypersurface-deformation
brackets in full generality, off-shell and with phase-space dependent gauge
functions, has revealed several new features of importance for modified and
quantized gravity in canonical form. In broad terms, there are three main
results, given by (i) an off-shell relationship between hypersurface
deformations and space-time Lie derivatives (Section~\ref{sec:Geometrodynamics
  revised}), (ii) the identification of differences between the canonical
Poisson brackets of gravitational constraints and the geometrical notion of
hypersurface deformations for phase-space dependent gauge functions
(Section~\ref{s:Offshell}), and (iii) an extension of the equations underlying
emergent modified gravity from spherically symmetric models to unrestricted
configurations (Section~\ref{s:new}). Taken together, these results imply that
a complete form of the canonical brackets of constraints, necessarily with
phase-space dependent gauge functions because of the presence of structure
functions, does not represent a single gravitational theory but rather a
larger framework in which different sectors of inequivalent space-time
theories can be embedded. In terms of geometrical hypersurface deformations,
the space-time structure of a single sector is determined by phase-space
independent lapse functions and shift vectors, and we can move between sectors
by considering linear combinations of the constraints with suitable
phase-space dependent lapse functions and shift vectors. Emergent modified
gravity consists precisely in the identification of such sectors, including
also other consistent modifications of the constraints.

Covariance conditions are the main part of our third broad result,
generalized here from spherically symmetric models to unrestricted space-time
configurations. These equations remain to be solved in order to see whether
new theories of emergent modified gravity are possible without symmetry
restrictions. However, the crucial equations remain valid almost unchanged,
suggesting that the phenomena of emergent modified gravity may be general. Our
second broad result, the identification of differences between canonical
constraint brackets and geometrical hypersurface deformations, explains why
such modifications of classical gravity may be possible: Within the vast set
of phase-space dependent constraint brackets, there may be independent
realizations of geometrical hypersurface deformations generated by phase-space
independent multipliers $\tilde{N}$ and $\tilde{N}^a$ of different fixed
phase-space dependent seed functions $N$ and $N^a$.

There are additional implications with connections to different
approaches. For instance, if the covariance condition (\ref{eq:Covariance
  condition - modified - reduced}) is extended off-shell, it is stronger than
the off-shell vanishing of the Jacobiator: If the covariance condition is
satisfied off-shell, the Jacobiator vanishes as readily seen from
(\ref{eq:Off-shell Jacobiator of three H}).  This property then ensures that
both the spatial metric via (\ref{eq:Covariance condition - modified -
  reduced}) as well as lapse and shift via (\ref{eq:Deformation of generators
  - canonical - Phase space dependent generators}) have off-shell
transformations as required for general covariance, their gauge
transformations corresponding to standard coordinate transformations.
However, requiring the conditions to hold only on-shell seems more natural
since this is the home of all physical processes, at least in classical-type
theories.  On-shell one can implement phase-space dependent gauge functions
into the covariance condition (\ref{eq:Covariance condition - modified -
  reduced}) quite trivially, while the Jacobiator (\ref{eq:Off-shell
  Jacobiator of three H}) always has to vanish on-shell even if the covariance
condition is not imposed. In particular, our results show that the
construction of anomaly-free (but not necessarily covariant) canonical
theories with generic coefficient functions in the Hamiltonian constraint
requires non-trivial conditions from the Jacobi identity. In the present
context, this requirement then rules out higher-derivative terms of the
momenta in anomaly-free Hamiltonian constraints, which in \cite{HigherCov} was
achieved by invoking the covariance condition.

Our result (\ref{eq:Deformation of generators - canonical - Phase space
  dependent generators}) ensures that the equations of motion, generated by
the same constraints as gauge transformations, are gauge covariant. Together
with (\ref{eq:Covariance condition - modified - reduced}), which makes sure
that gauge transformations are equivalent to coordinate transformations, the
equations of motions are generally covariant.  Any theory of this form is
therefore generally covariant according to the usual understanding.
Nevertheless, our general formulations allow for equations of motion that are
not necessarily derivable from an invariant action where all space-time
tensors are suitably contracted and the space-time metric is used in the
volume element.  The geometric and covariance conditions derived here are
therefore weaker than demanding the existence of an invariant action
functional. They lead to a larger class of generally covariant equations
of motion.

\section*{Acknowledgements}

We thank Madhavan Varadarajan for detailed discussions and suggestions, as
well as his hospitality at RRI. This work was supported in part
by NSF grant PHY-2206591.

\begin{appendix}

  \section{Decomposition of Lie derivatives}
  \label{a:ADM}
  
Given a globally hyperbolic space-time,
  $M = \Sigma \times \mathbb{R}$, with dimension ${\rm dim} (M) = d$, the 
  space-time metric in ADM form is given by
\begin{eqnarray}
    {\rm d} s^2 = \sigma N^2 {\rm d} t^2 + q_{a b} ( {\rm d} x^a + N^a {\rm d}
  t ) ( {\rm d} x^b + N^b {\rm d} t )
    \,.
    \label{eq:ADM line element - Geometrodynamics}
\end{eqnarray}
The coordinate frame refers to an observer with time-evolution vector field 
\begin{equation}
    t^\mu = N n^\mu + N^a s_a^\mu
    \ .
    \label{eq:Time-evolution vector field2}
\end{equation}
Here, $q_{a b} (t)$ is the family of spatial metrics on the $d-1$ dimensional
submanifolds $\Sigma_t$ that foliate $M$.  The inverse space-time metric equals
\begin{eqnarray}
    g^{\mu \nu} =
    q^{a b} s^\mu_a s^\nu_a
    + \frac{\sigma}{N^2} \left(t^\mu - N^a s^\mu_a\right) \left(t^\nu - N^b s^\nu_b\right)
    \ .
\end{eqnarray}

In the main text, we used several components of the ADM decomposition of
the Lie derivative $\mathcal{L}_{\xi} g_{\mu \nu} = \xi^\alpha \partial_\alpha g_{\mu \nu} +
g_{\mu \alpha} \partial_\nu \xi^\alpha + g_{\nu \alpha} \partial_\mu
\xi^\alpha$ along a space-time vector field $\xi^{\mu}$ related to components $\epsilon^{\mu}$
in a normal frame by (\ref{eq:Diffeomorphism generator projection}). In explicit form,
they are:
\begin{eqnarray}
    \mathcal{L}_\xi g_{a b}
    &=&
    \xi^t \partial_\alpha q_{a b}
    + g_{a t} \partial_b \xi^t
    + g_{b t} \partial_a \xi^t
    + \xi^c \partial_c q_{a b}
    + q_{a c} \partial_b \xi^c
    + q_{b c} \partial_a \xi^c
    \nonumber\\
    &=&
    \frac{\epsilon^0}{N} \partial_\alpha q_{a b}
    + g_{a t} \partial_b \left(\frac{\epsilon^0}{N}\right)
    + g_{b t} \partial_a \left(\frac{\epsilon^0}{N}\right)
    \nonumber\\
    &&+ \left( \epsilon^c - \frac{\epsilon^0}{N} N^c \right) \partial_c q_{a b}
    + q_{a c} \partial_b \left( \epsilon^c - \frac{\epsilon^0}{N} N^c \right)
    + q_{b c} \partial_a \left( \epsilon^c - \frac{\epsilon^0}{N} N^c \right)
    \nonumber\\
    &=&
    \frac{\epsilon^0}{N} \partial_\alpha q_{a b}
    + N^c q_{c a} \partial_b \left(\frac{\epsilon^0}{N}\right)
    + N^c q_{c b} \partial_a \left(\frac{\epsilon^0}{N}\right)
    \nonumber\\
    &&+ \left( \epsilon^c - \frac{\epsilon^0}{N} N^c \right) \partial_c q_{a b}
    + q_{a c} \partial_b \left( \epsilon^c - \frac{\epsilon^0}{N} N^c \right)
    + q_{b c} \partial_a \left( \epsilon^c - \frac{\epsilon^0}{N} N^c \right)
    \nonumber\\
    &=&
    \frac{\epsilon^0}{N} \dot{q}_{a b}
    + \left( \epsilon^c - \frac{\epsilon^0}{N} N^c \right) \partial_c q_{a b}
    + q_{a c} \left( \partial_b \epsilon^c - \frac{\epsilon^0}{N} \partial_b N^c \right)
    + q_{b c} \left( \partial_a \epsilon^c - \frac{\epsilon^0}{N} \partial_a N^c \right)
    \nonumber\\
    &=&
    \frac{\epsilon^0}{N} \dot{q}_{a b}
    + \epsilon^c \partial_c q_{a b}
    + q_{c a} \partial_{b} \epsilon^c
    + q_{c b} \partial_{a} \epsilon^c
    - \frac{\epsilon^0}{N} \left( N^c \partial_c q_{a b}
    + q_{c a} \partial_{b} N^c
    + q_{c b} \partial_{a} N^c \right)
\end{eqnarray}
for the spatial part,
\begin{eqnarray}
    \mathcal{L}_\xi g_{t t}
    &=&
    \xi^\alpha \partial_\alpha g_{t t}
    + 2 g_{t \alpha} \partial_t \xi^\alpha
    \nonumber\\
    &=&
    \xi^t \partial_t g_{t t}
    + 2 g_{t t} \partial_t \xi^t
    + \xi^c \partial_c g_{t t}
    + 2 g_{t c} \partial_t \xi^c
    \nonumber\\
    &=&
    \frac{\epsilon^0}{N} \partial_t g_{t t}
    + 2 g_{t t} \partial_t \left(\frac{\epsilon^0}{N}\right)
    + \left(\epsilon^c - \frac{\epsilon^0}{N} N^c\right) \partial_c g_{t t}
    + 2 g_{t c} \partial_t \left(\epsilon^c - \frac{\epsilon^0}{N} N^c\right)
    \nonumber\\
    &=&
    \frac{\epsilon^0}{N} \partial_t \left( \sigma N^2 + N^c N^d q_{c d} \right)
    + 2 \left( \sigma N^2 + N^c N^d q_{c d} \right) \partial_t \left(\frac{\epsilon^0}{N}\right)
    \nonumber\\
    &&+ \left(\epsilon^e - \frac{\epsilon^0}{N} N^e\right) \partial_e \left( \sigma N^2 + N^c N^d q_{c d} \right)
    + 2 N^d q_{d c} \partial_t \left(\epsilon^c - \frac{\epsilon^0}{N} N^c\right)
    \nonumber\\
    &=&
    2 \sigma N \dot{\epsilon}^0
    + N^c N^d \frac{\epsilon^0}{N} \dot{q}_{c d}
    \nonumber\\
    &&
    + \left(\epsilon^e - \frac{\epsilon^0}{N} N^e\right) \left( \sigma 2 N \partial_e N + N^c N^d \partial_e q_{c d} + 2 N^d q_{c d} \partial_e N^c \right)
    + 2 N^d q_{d c} \dot{\epsilon}^c
    \nonumber\\
    &=&
    \sigma 2 N \left( \dot{\epsilon}^0
    + \epsilon^a \partial_a N
    - N^a \partial_a \epsilon^0 \right)
    + N^a N^b \mathcal{L}_\xi g_{a b}
    \nonumber\\
    &&+ 2 q_{a b} N^a \left(\dot{\epsilon}^b + \epsilon^c \partial_c N^b - N^c \partial_c \epsilon^b
    - \sigma q^{b c} \left(\epsilon^0 \partial_c N - N \partial_c \epsilon^0 \right)\right)
    \nonumber\\
    &=&
    \sigma 2 N \delta_\epsilon N
    + N^a N^b \mathcal{L}_\xi g_{a b}
    + 2 q_{a b} N^a \delta_\epsilon N^b
\end{eqnarray}
for the time component (related to the lapse function), and
\begin{eqnarray}
    \mathcal{L}_\xi g_{t a}
    &=&
    \xi^\alpha \partial_\alpha g_{t a}
    + g_{t \alpha} \partial_a \xi^\alpha
    + g_{a \alpha} \partial_t \xi^\alpha
    \nonumber\\
    &=&
    \xi^t \partial_t g_{t a}
    + g_{t t} \partial_a \xi^t
    + g_{a t} \partial_t \xi^t
    + \xi^b \partial_b g_{t a}
    + g_{t b} \partial_a \xi^b
    + q_{a b} \partial_t \xi^b
    \nonumber\\
    &=&
    \frac{\epsilon^0}{N} \partial_t g_{t a}
    + g_{t t} \partial_a \left(\frac{\epsilon^0}{N}\right)
    + g_{a t} \partial_t \left(\frac{\epsilon^0}{N}\right)
    + \left( \epsilon^b - \frac{\epsilon^0}{N} N^b\right) \partial_b g_{t a}\nonumber\\
&&    + g_{t b} \partial_a \left( \epsilon^b - \frac{\epsilon^0}{N} N^b\right)
    + q_{a b} \partial_t \left( \epsilon^b - \frac{\epsilon^0}{N} N^b\right)
    \nonumber\\
    &=&
    \frac{\epsilon^0}{N} \partial_t \left(N^b q_{b a}\right)
    + \left(\sigma N^2 + N^c N^d q_{c d}\right) \partial_a \left(\frac{\epsilon^0}{N}\right)
    + N^b q_{b a} \partial_t \left(\frac{\epsilon^0}{N}\right)
    \nonumber\\
    &&    + \left( \epsilon^b - \frac{\epsilon^0}{N} N^b\right) \partial_b \left( N^c q_{c a}\right)
+ N^c q_{c b} \partial_a \left( \epsilon^b - \frac{\epsilon^0}{N} N^b\right)
    + q_{a b} \partial_t \left( \epsilon^b - \frac{\epsilon^0}{N} N^b\right)
    \nonumber\\
    &=&
    \frac{\epsilon^0}{N} N^b \dot{q}_{b a}
    + \sigma N \partial_a \epsilon^0
    - \sigma \epsilon^0 \partial_a N
    + \epsilon^b q_{c a} \partial_b N^c
    - \frac{\epsilon^0}{N} N^b q_{c a} \partial_b N^c\nonumber\\
&&    + \epsilon^b N^c \partial_b q_{c a}
    - \frac{\epsilon^0}{N} N^b N^c \partial_b q_{c a}
    \nonumber\\
    &=&
    N^b \mathcal{L}_\xi g_{b a}
    + q_{a b} \left( \dot{\epsilon}^b
    + \epsilon^c \partial_c N^b
    - N^c \partial_{c} \epsilon^b
    - \sigma q^{b c} \left( \epsilon^0 \partial_c N - N \partial_c \epsilon^0 \right) \right)
    \nonumber\\
    &=&
    N^b \mathcal{L}_\xi g_{b a}
    + q_{a b} \delta_\epsilon N^b\label{eq:ADM decomposition of Lie derivative of metric - Geometrodynamics}
\end{eqnarray}
for the time-space part (related to the shift vector).
Similarly, the inverse metric components have the Lie derivatives
\begin{eqnarray}
    \mathcal{L}_\xi g^{t \mu}
    &=& 
    \xi^\alpha \partial_\alpha g^{t \mu}
    - g^{\alpha \mu} \partial_\alpha \xi^t
    - g^{t \alpha} \partial_\alpha \xi^\mu
    \nonumber\\
    &=&
    \xi^t \partial_t g^{t \mu}
    - g^{t \mu} \partial_t \xi^t
    - g^{t t} \partial_t \xi^\mu
    + \xi^c \partial_c g^{t \mu}
    - g^{c \mu} \partial_c \xi^t
    - g^{t c} \partial_c \xi^\mu
    \nonumber\\
    &=&
    \frac{\epsilon^0}{N} \partial_t g^{t \mu}
    - g^{t \mu} \partial_t \left(\frac{\epsilon^0}{N}\right)
    - g^{c \mu} \partial_c \left(\frac{\epsilon^0}{N}\right)
    \nonumber\\
    &&- \frac{\sigma}{N^2} \partial_t \xi^\mu
    + \left( \epsilon^c - \frac{\epsilon^0}{N} N^c \right) \partial_c g^{t \mu}
    + \frac{\sigma}{N^2} N^c \partial_c \xi^\mu
\end{eqnarray}
as well as
\begin{eqnarray}
    \mathcal{L}_\xi g^{t t}
    &=&
    - \sigma \frac{2}{N^3}
    \left( \dot{\epsilon}^0
    + \epsilon^c \partial_c N 
    - N^c \partial_c \epsilon^0
    \right)
    = - \sigma \frac{2}{N^3} \delta_\epsilon N
\end{eqnarray}
and
\begin{eqnarray}
    \mathcal{L}_\xi g^{t a}
    &=&
    - \sigma \frac{1}{N^2} \left( \delta_\epsilon N^a
    - \frac{2}{N} N^a \delta_\epsilon N \right)
    \ .
\end{eqnarray}

\section{Poisson brackets for linear combinations of constraints}
\label{a:Brackets}

Using phase-space dependent coefficients $B$ (a spatial scalar without density weight)
and $A^a$ (a spatial vector field) for linear combinations
of the classical constraints, the Poisson bracket of a Hamiltonian constraint
with lapse function $B\tilde{N}$ and the diffeomorphism constraint is given by
\begin{eqnarray}
&&    \{ H [ B \tilde{N}] , \tilde{H}_a [\tilde{M}^a] \}\nonumber\\
    &=&
    \int {\rm d}^3 x {\rm d}^3 y\ \left(  
    \{ H (x) , H_a (y) \} B (x) \tilde{N} (x) \tilde{M}^a (y)
    + \{ B (x) , \tilde{H}_a (y) \} H (x) \tilde{N} (x) \tilde{M}^a (y)
    \right)
    \nonumber\\
    &=&
    - H \left[ \tilde{M}^a \partial_a \left( B \tilde{N} \right) \right]
    + H [ \tilde{M}^a (\partial_a B) \tilde{N}]
  =
    - H \left[ B \tilde{M}^a \partial_a \tilde{N} \right]\,.
\end{eqnarray}
The Poisson bracket of two diffeomorphism constraints, one with $A^b\tilde{N}$
and one with a generic shift vector, is given by
\begin{eqnarray}
&&    \{ H_b [ A^b \tilde{N} ] , \tilde{H}_a [\tilde{M}^a] \}\nonumber\\
    &=&
    \int {\rm d}^3 x {\rm d}^3 y\ \left(
    \{ H_b (x) , \tilde{H}_a (y) \} A^b (x) \tilde{N} (x) \tilde{M}^a (y)
    + \{ A^b (x) , \tilde{H}_a (y) \} H_b (x) \tilde{N} (x) \tilde{M}^a (y)
    \right)
    \nonumber\\
    &=&
    - H_a \left[\mathcal{L}_{\vec{\tilde{M}}} \left( A^a \tilde{N} \right)\right]
    + H_b [( \mathcal{L}_{\vec{\tilde{M}}} A^b) \tilde{N} ]
  =
    - H_a \left[ A^a \tilde{M}^b \partial_b \tilde{N} \right]\,.
\end{eqnarray}
The Poisson bracket of two Hamiltonian constraints with lapse functions
multiplied by $B$ equals
\begin{eqnarray}
&&    \{ H [ B \tilde{N} ] , H [ B \tilde{M} ] \}\nonumber\\
    &=&
    \int {\rm d}^3 x {\rm d}^3 y\ \bigg( 
    \{ H (x) , H (y) \} B (x) \tilde{N} (x) B (y) \tilde{M} (y)
    + \{ H (x) , B (y) \} B (x) \tilde{N} (x) H (y) \tilde{M} (y)
    \nonumber\\
    &&
    + H (x) \{ B (x) , H (y) \} \tilde{N} (x) B (y) \tilde{M} (y)
    + H (x) \{ B (x) , B (y) \} \tilde{N} (x) H (y) \tilde{M} (y)
    \bigg)
    \nonumber\\
    &=&
    H_a \left[ \sigma q^{a b} \left( (B \tilde{M}) \partial_b (B \tilde{N}) - (B\tilde{N}) \partial_b (B \tilde{M}) \right)\right]
    \nonumber\\
    &&
    + \int {\rm d}^3 x {\rm d}^3 y\ H (x) \bigg( 
    \{ B (x) , H (y) \} \left( \tilde{N} (x) B (y) \tilde{M} (y) - \tilde{M} (x) B (y) \tilde{N} (y) \right)
    \nonumber\\
    &&
    + \{ B (x) , B (y) \} \tilde{N} (x) H (y) \tilde{M} (y)
    \bigg)\,,
\end{eqnarray}
and the Hamiltonian constraint with a lapse function multiplied by $B$
together with a diffeomorphism constraint with a shift vector obtained from
a multiple of $A^a$ have the Poisson bracket
\begin{eqnarray}
&&    \{ H [ B \tilde{N} ] , H_a [A^a \tilde{M}] \}\nonumber\\
    &=&
    \int {\rm d}^3 x {\rm d}^3 y\ \bigg(
    \{ H (x) , H_a (y) \} B (x) \tilde{N} (x) A^a (y) \tilde{M} (y)
    \nonumber\\
    &&
    + \{ H (x) , A^a (y) \} B (x) \tilde{N} (x) H_a (y) \tilde{M} (y)
    + \{ B (x) , H_a (y) \} H (x) \tilde{N} (x) A^a (y) \tilde{M} (y)
    \nonumber\\
    &&
    + \{ B (x) , A^a (y) \} H (x) \tilde{N} (x) H_a (y) \tilde{M} (y)
    \bigg)
    \nonumber\\
    &=&
    - H \left[ (A^a \tilde{M}) \partial_b (B \tilde{N}) \right]
    + H \left[ (A^a \tilde{M}) \partial_a B \tilde{N} \right]
    \nonumber\\
    &&
    + \int {\rm d}^3 x {\rm d}^3 y\ \bigg(
    - \{ A^a (y) , H (x) \} B (x) \tilde{N} (x) H_a (y) \tilde{M} (y)
    \nonumber\\
    &&
    + \{ B (x) , A^a (y) \} H (x) \tilde{N} (x) H_a (y) \tilde{M} (y)
    \bigg)
    \nonumber\\
    &=&
    - H \left[ B A^a \tilde{M} \partial_b \tilde{N} \right]
    + \int {\rm d}^3 x {\rm d}^3 y\ \bigg(
    - \{ A^a (y) , H (x) \} B (x) \tilde{N} (x) H_a (y) \tilde{M} (y)
    \nonumber\\
    &&
    + \{ B (x) , A^a (y) \} H (x) \tilde{N} (x) H_a (y) \tilde{M} (y)
    \bigg)\,.
\end{eqnarray}
Finally, for two diffeomorphism constraints with shift vectors given by
multiples of $A^a$, we obtain the Poisson bracket
\begin{eqnarray}
&&    \{ H_a [A^a \tilde{N}] , H_b [A^b \tilde{M}] \}\nonumber\\
    &=&
    \int {\rm d}^3 x {\rm d}^3 y\ \bigg(
    \{ H_a (x) , H_b (y) \} A^a (x) \tilde{N} (x) A^b (y) \tilde{M} (y)
    \nonumber\\
    &&
    + \{ H_a (x) , A^b (y) \} A^a (x) \tilde{N} (x) H_b (y) \tilde{M} (y)
    \nonumber\\
    &&
    + \{ A^a (x) , H_b (y) \} H_a (x) \tilde{N} (x) A^b (y) \tilde{M} (y)
    + \{ A^a (x) , A^b (y) \} H_a (x) \tilde{N} (x) H_b (y) \tilde{M} (y)
    \bigg)
    \nonumber\\
    &=&
    H_b \left[\mathcal{L}_{\vec{A} \tilde{N}} (A^b \tilde{M})\right]
    - H_b[ \tilde{M} \mathcal{L}_{\vec{A} \tilde{N}} A^b ]
    + H_b[ \tilde{N} \mathcal{L}_{\vec{A} \tilde{M}} A^b ]
    \nonumber\\
    &&
    + \int {\rm d}^3 x {\rm d}^3 y\ \{ A^a (x) , A^b (y) \} H_a (x) \tilde{N} (x) H_b (y) \tilde{M} (y)
    \nonumber\\
    &=&
    \int {\rm d}^3 x {\rm d}^3 y\ \{ A^a (x) , A^b (y) \} H_a (x) \tilde{N} (x) H_b (y) \tilde{M} (y)
    \ .
\end{eqnarray}
These individual Poisson brackets can be combined to
obtain the full constraint brackets for linear combinations as described in
the main text.
  
\end{appendix}


\end{document}